\documentclass[10pt]{article}

\usepackage{graphicx}

\newtheorem{theorem}{Theorem}[section]

\begin{document}
\begin{center}
{\Large RESONANCE QUANTUM SWITCH :
SEARCH FOR \\
WORKING PARAMETERS}\\
\vskip1cm {\large N.T.\,Bagraev $^1$, A.B.\,Mikhailova $^2$,\\[0pt]
B.S.\,Pavlov $^{2,3}$, L.V.\,Prokhorov $^2$, A.M.\,Yafyasov $^2$}\\
\end{center}
\vskip10pt \noindent $^1$ A.F. Ioffe Physico-Technical Institute,
St. Petersburg, 194021, Russia.\newline $^2$ V.A. Fock Institute
of Physics, St.Petersburg State University,St.Petersburg, 198504,
Russia\newline $^3$Department of Mathematics , University of
Auckland, Private Bag 92019, Auckland, New Zealand;

\vskip10pt

\centerline{\large\bf Abstract}{\noindent \bf Design of a
three-terminal Quantum Switch is suggested in form of  a network
consisting of a circular quantum well and four semi-infinite
single mode quantum wires attached to it. It is shown that in
resonance case, when the Fermi level in the wires is  close to
some  energy level in the well, the magnitude of the governing
electric field on the  well may be specified in such a way that
the quantum current across the switch from the up-leading wire to
the outgoing wires (terminals) can be controlled via rotation of
the orthogonal projection of the governing electric field onto the
plane of the device. The details of design of the switch are
chosen in dependence of desired working temperature, available
Fermi level and effective mass of electron. The speed  of
switching is estimated. A solvable model of the  switch in form of
a one-dimensional graph with a resonance vertex is suggested.}
\vskip5pt

  \vskip10pt
  \noindent
PACS numbers: 73.63.Hs,73.23.Ad,85.35.-p,85.35.Be

\noindent
 Keywords: Quantum well, Single-mode transport, Resonance
Scattering
\section{Introduction}
Basic problems of quantum conductance were related to Scattering
Processes long ago, see \cite{Landauer70,Buttiker85,PalmThil92}
and the role of scattering in mathematical design of quantum
electronic devices was  clearly understood  by the beginning  of
nineties, see \cite{Adamyan,Buttiker93}. Still the practical
design of devices, beginning from the classical Esaki diode
\cite{Esaki}  up to modern types, see for instance \cite{Compano},
was based on the resonance of energy levels rather than on
resonance properties of the corresponding wave functions.
Importance of  the interference in mathematical design of  devices
was noticed in \cite{Exner88,Alamo} and intensely  studied in
\cite{PalmThil92,Exner96,Interf1,Interf2}, see also recent papers
\cite{Safi99,Kouw2002}.
\par
  Modern experimental technique already permits to observe
resonance effects caused by details of the shape of the resonance
wave functions, see \cite{B1,B2,B3,Carbon01}. We propose using  of
these effects as a natural tool for manipulation of the electron's
current in the  quantum switch with one input  wire  and  three
terminals.
\par
Our problem on mathematical design of a three-terminal Quantum
Switch RQS-3 (for triadic logic) appeared first as a Work-Package
in the ES-Project (joint with Solvay Institute) "New technologies
for narrow-gap semiconductors" ESPRIT-28890 NTCONGS, 1998 - 1999.
Results of the project were initially formulated in rather
mathematical language in \cite {MP00}, for  the  switch based  on
the  quantum ring, in  \cite{MP01}, and later in  \cite{MP02} for
the  switch based on the  quantum well and patented in \cite{P00}.
Prospects of practical implementation of the devices were
discussed in \cite{P01,Helsinki2}. Note, that an attempt of
computation of the quantum current through the Y-junction was also
done in recent paper \cite{CsXu02}, but an important  connection
between  the geometric characteristics of the  junction, encoded
in the corresponding Dirichlet-to-Neumann map, was not noticed
there.
\par
Our initial mathematical idea of the resonance manipulation of the
{\it single-mode} quantum current was based on an observation from
\cite{Opening}: \vskip0.5cm {\it The resonance transmission across
the quantum system caused by attachment of incoming and outgoing
channels  is proportional to the products of proper local
characteristics of the corresponding resonance eigenfunction of
the  system at the places where the channels are attached.}
\vskip0.5cm Some mathematical aspects of this observation were
discussed in papers \cite{MP01,MP02,P02,QSW}. In actual paper we
describe a parameter regime where the quantum switch works. The
Resonance Quantum Switch is  formed as a Quantum network $\Omega$
on the surface $R_{_2}$ of a semiconductor  and consists of a deep
quantum well $\Omega_{_{0}}$ and few equivalent quantum wires
$\Omega_{_{1}},\, \Omega_{_{2}},\, \Omega_{_{3}}\, \Omega_{_{4}}$
attached to it .
\par
The role of Hamiltonian of the corresponding quantum system is
played by the one-electron Schr\"{o}dinger operator $l$, see below
in section 2. The potential is  constant  $V_{_{\infty}}$ in the
wires, linear (corresponding to  the macroscopic electric  field )
on the well and  zero on the complement $R_{_2}\backslash \Omega :
= \Omega'$ of the  network. If the depth of the Fermi level in the
wires with respect to the potential on the complement of the
network is large enough, see estimates below in sub-section 2.1,
one can replace the matching boundary conditions on the border of
the network  with the complementary domain $\Omega'$ by the
homogeneous Dirichlet conditions, at least for electrons with
energy close to  the Fermi level.
\begin{figure}[ht]
  \begin{center}
  \includegraphics{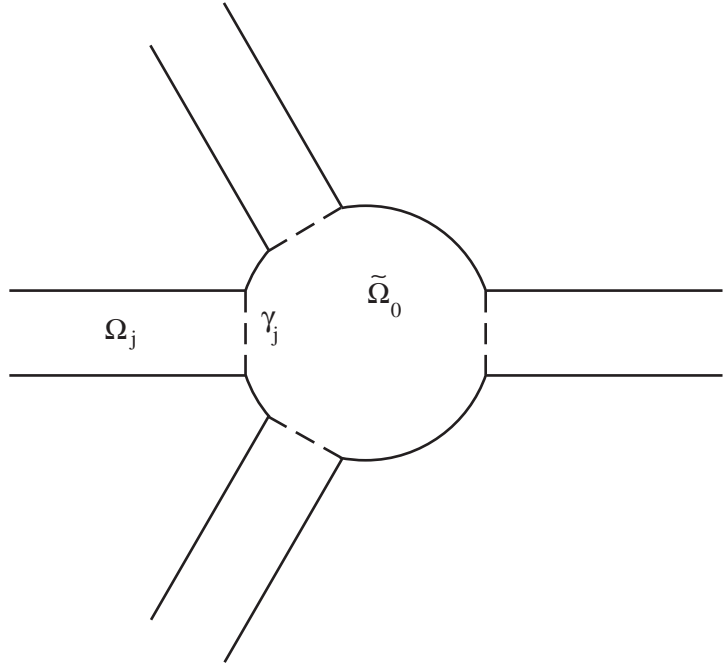}
  \end{center}
  \caption{Resonance Quantum Switch. The circular quantum well $\Omega_0$ with wires
   $\Omega_j,\, j=1,2,3,4$ attached. The modified well $\widetilde{\Omega}_0$ has small
  circular arcs of the circular boundary replaced by the flat bottom sections
  ${\bf\gamma_{_j}}$ of the  wires.}
  \label{F:figure}
\end{figure}
 We do not assume
actually that the wires are thin (see the discussion below in the
subsection (2.3) or connection of the wires to the quantum well is
weak, but we assume that the dynamics of electrons in the wires is
single-mode and ballistic on large intervals of the wires,
compared  with the size of the geometric details of the
construction ( the width of the wire or the size of the contacts).
 We  consider the  electron's transmission  across the
well from one  wire to another  as  a Scattering process. It
appears that the transmission coefficients   can be
 manipulated  via  macroscopical electric  field applied  to the
 quantum well, see section 5. We aim at the estimation of the temperature
stability of the device and the speed of switching in dependence
on geometrical details of the construction and properties of
selected materials.
\par
Note that it is  impossible to construct a solution to the
Schr\"{o}dinger equation on the Quantum network in analytic form.
On the  other hand the direct computing is non efficient for
optimization of  the construction because of large number of
essential parameters and the resonance  switching effect observed
only  on the  triple (or multiple) point  in the space of
essential physical
 and geometrical parameters.
\par
Analysis  of the one-dimensional Schr\"{o}dinger equation on the
corresponding one-dimensional graph  is a comparatively  simple
alternative of explicit formulae for the solution of the  above
Scattering problem, but estimation of errors appearing  from the
substitution of the Quantum network- the  ``fattened graph"- by
the corresponding one-dimensional graph may be difficult. In
mathematical papers \cite{KZ01,RS01} the authors study  the
spectrum of the Schr\"{o}dinger operator on a compact ``fattened
graph". Based on variational approach developed in \cite{Schat96}
they noticed that the (discrete) spectrum of Laplacian on a system
of finite length shrinking wave-guides width $\delta$ attached to
the shrinking vertex domain diameter $R =R_{_{1}}\,
\delta^{\alpha},\, 0< \alpha < 1 $, tends to the spectrum of
Laplacian  on the corresponding one-dimensional graphs, but with
different boundary conditions at vertices, depending on the speed
of shrinking.
\par
In distinction from the papers \cite{KZ01,RS01} quoted above, our
approach to the  Schr\"{o}dinger equation on the ``fattened
graph'', see \cite{MP02,P02},  is based on analysis of the
resonance transmission through the quantum well radius $R$ with
few quantum wires width $\delta$ attached to it. Assuming that
Fermi level is situated  on the  first spectral band  in the
wires, we  impose additional ``chopping off''  boundary conditions
on the bottom sections of the semi-infinite wires. These condition
split the original operator  into  orthogonal sum of the  trivial
part $\left\{\sum_{_{s = 1}}^{^4} l^{^r}_{_{s}}\right\}$ in the
open channels and a non-trivial part $l^{^r}_{_{0}}$ which  plays
the role  of an ``Intermediate operator": $ l \to  \left\{
\sum_{_{s = 1}}^{^4} l^{^r}_{_{s}} \right\} \oplus l^{^r}_{_{0}}.
$ The  spectrum  of the  intermediate  operator consists of an
absolutely continuous part of varying  multiplicity which  begins
from the second threshold in the wires and a sequence of
eigenvalues $\lambda^{^r}_{_{l}}$ which can accumulate at
infinity. The eigenvalues  which  are sitting on the first
spectral band generically give rise to resonances of the
Scattering problem on the  network and hence define the resonance
conductance. Our main tool is the following approximate formula
for  the Scattering matrix in terms of eigenfunctions
$\varphi_{_l}$ of continuous and  discrete spectrum of the
Intermediate operator and  corresponding eigenvalues:
\begin{equation}
\label{Sapprox} {\bf S} (\lambda) \approx \frac{i{\bf p}(\lambda)I
+ {\cal D}{\cal N}_{_{T}}}{i{\bf p}(\lambda)I - {\cal D}{\cal
N}_{_{T}}} := {\bf S}_{_{T}} (\lambda),
\end{equation}
where  $i\,{\bf p}\,(\lambda)$ is, in simplest case, an exponent
from the bounded modes $\displaystyle e^{^{\pm i\,{\bf
p}\,(\lambda) x}} e_{_s}$ in  open channels, spanned by the
corresponding cross-section eigenfunction $e_{_s}$, and
\begin{equation}
\label{DNess} {\cal D}{\cal N}^{^r} (\lambda)_{_T} = \sum_{_{l}}
\frac{P_{_+}\frac{\partial \varphi^{^r} _{_l}}{\partial n}\rangle
\langle P_{_+}\frac{\partial \varphi^{^r} _{_l}}{\partial
n}}{\lambda - \lambda^{^r}_l} + \int \frac{P_{_+}\frac{\partial
\varphi^{^r} _{_l}}{\partial n}\rangle \langle
P_{_+}\frac{\partial \varphi^{^r} _{_l}}{\partial n}}{\lambda^{^r}
- \lambda_{_l}} d\rho(\lambda_{_l})
\end{equation}
is an essential part of  the  Dirichlet-to-Neumann  map of the
Intermediate  operator $l_{_0}^{^r} $, which contains  summation
and integration over  spectrum   $\left\{\lambda_{l}\right\}$ of
the Intermediate operator situated  on  the essential interval of
energy $|\frac{\hbar^{^2}}{2 m_{_0}}\,\,\lambda_{_{l}} - E_{_{F}}|
< \kappa T$ for given temperature.  The operator $P_{_+}$ is a
projection $\sum_{_{s=1}}^{^{4}} e_{_s}\rangle\, \langle e_{_s}$
onto the subspace $E_{_+} = \bigvee_{_{s=1}}^{^{4}} e_{_s}$
spanned  by the cross-section eigen-functions of  open channels
on the bottom sections of the wires.
\par
In most interesting case when only one resonance eigenvalue
$\lambda^{^r}_{_0}$ is sitting on the essential spectral band, the
above formula (\ref{Sapprox}) gives important ``one-pole
approximation'' which is  used  below for estimation of the
working parameters of the switch.
\par
Our paper has the following plan. In the second  section we
describe  the  Hamiltonian and  supply  (in 2.2) some  physical
motivation supporting the  choice of basic  parameters and
generally motivate our  approach. In the third section we
formulate the relevant Scattering Problem and announce the
formula for the Scattering matrix in terms of the
Dirichlet-to-Neumann map (DN-map) of the Intermediate operator.
The derivation of this  formula  is postponed  to  Appendix,
sub-section 10.2. In fourth section  we  apply the  announced
formula
 for  estimation of the  speed of switching, and derive from it a
 useful  approximate  formula  which is  valid for  relatively
 low  temperature - the ``one-pole approximation''
of the  Scattering matrix.
 In fifth sections we discuss the resonance switching
phenomenon. In the section  6 we suggest a procedure of choosing
of the working point of the device in dependence on desired
working temperature and in  the  last  section 7 we  describe  a
solvable  model  of the  switch  in form of a  one-dimensional
graph with a  resonance vertex.
 In Appendix we supply technical
details on DN-map, derive  the  announced formula for the
Scattering matrix and interpret  the  corresponding one-pole
approximation (\ref{onepole})  as a Scattering matrix  of some
solvable  model.
\section{Design of the  switch}
Here we  provide  a  preliminary  discussion of  physical and
geometrical limitations  of  the  switch. A  basic example is
considered in sub-section 2.3.
 \subsection{Schr\"{o}dinger equation}
Conductance of the network constructed  on the  surface of the
crystallized  medium  depends on geometry of the network and on
the correspondence between the crystal structure and the form and
positions of the wires with respect to the  geometry of  the
crystal's lattice . In simplest case this  dependence is encoded
in tensor of effective mass of the electron and  in basic
potentials in the wires and on the  well. For instance, on
$(100)$-plane in the Si matrix there are two lower and four upper
valleys. For lower valleys effective masses are equal $m^{\bot}=
m_t= 0.190\,\, m_0$ across the valley and $m^{\parallel}= m_e =
0.916\,\, m_0$ along the valley, where $m_0$ is  the conventional
electron mass. On the plane  $(110)$ there are  four lower valleys
with effective masses calculated as
 $\frac{m_t + m_e}{2} = 0.553\,\, m_0,\, \frac{2 m_t\, m_e}{m_t + m_e}= 0.315 m_0$
respectively and  two  upper valleys  with  masses $m_t,\,
m_e,\,m_t$. For  quantum wires formed on the  surface of some
narrow-gap semiconductors, like CdHgTe, the tensor of  effective
mass is isotropic, and the  value of the  effective  mass is
small, $m^{^{\parallel}} = m^{^{\bot}} << m_{_0}$, see \cite{Shur}
and the table in section 6. In following  sections we  assume that
in case of  switch based on Si the  effective  masses across and
along the  vires  are the  same for  all  wires: $m^{\bot} =
0.190\,\, m_0$ and $m^{\parallel} = 0.916\,\, m_0$ respectively.
Note  that  the resonance  switching phenomenon is defined  by the
geometry of the  well, see section 5, and  is invariant with
respect to the magnitudes of effective masses : the masses may be
different in different wires. Though the  transmission
coefficients depend on effective masses, they can be easily
calculated, once the  orientation of the wires with respect to the
crystal's structure and the  corresponding  effective masses are
known.
\par
The bottom of the  quantum well in a crystallized medium  is
generally a sophisticated network of valleys. We assume that  the
fine structure  of  this  pattern permits to replace  the
corresponding Schr\"{o}dinger equation with anisotropic effective
mass by  the standard  Schr\"{o}dinger equation with  an average
effective mass:
\begin{equation}
\label{Schredeq} - \frac{\hbar^2}{2 m}\bigtriangleup  +  V(x) = E
u.
\end{equation}
 \par
We assume that quantum dynamics outside the network is generated
by similar Schr\"{o}dinger equation with zero potential on the
complement of the network $V_{_{out}}(x)= 0,\,\, x \in \Omega' =
\in R_2 \backslash \Omega$, and the averaged effective mass for
both equations coincides with the conventional electron mass
$m_{_{0}}$.
\par
 If the Fermi level is deep enough counting from  the level of the
potential outside the network, compared with the radius of the
well:
\[
\frac{2m_0 (0 - E_{_{F}})}{\hbar^2} \, R^2 = 0.3136 \,\,(0 -
E_f)\,R^2 >> 1,
\]
($R$ measured in angstr\"{o}ms and the depth of the Fermi level
$(0 - E_{_F})$ in electron-volts), then for the values of energy
$E$ near to the Fermi-level $E_{_{F}}$ the spectral problem can be
reduced to the spectral problem for the corresponding
Schr\"{o}dinger operator  on the network $\Omega : =
\Omega_0\cup\Omega_1\cup \Omega_2\cup\Omega_{3}\cup\Omega_{4}$
with the mixed boundary condition on the boundary
$\partial\Omega$:
\[
\frac{\partial u}{\partial n} + \sqrt{\frac{2m (V_{_{out}} -
E_{_{F}})}{\hbar^2}}u \bigg|_{\partial \Omega} = 0,
\]
or even with zero (Dirichlet homogeneous) boundary condition, if
the above ratio is large enough. In this paper we just assume that
\[
u \big|_{\partial \Omega_0} = 0.
\]
 The potential $V(x)$ on the  quantum well is defined  by  the
 macroscopic``governing" electric  field ${\cal E}\,\nu$ which is constant inside
the well, $V(x) = {\cal E} e \langle x,\,\nu\rangle  + V_{_{0}}$,
where $e$ is the electron charge and the unit vector  $\nu $ shows
the direction of the field. The  magnitude of the  field is
specified in section 5.
\par
In the  wires  $\Omega_{_s}$: $-{\bf l}< x_{_{s}} < \infty,\,\, 0<
y <\delta$ the potential is piece-wise constant
\[
V_{_{s}}(x,y)= \left\{
\begin{array}{ccc}
V_{_{\infty}} + \frac{\hbar^{^2}\, H^{^2}}{2m_{_0}}  &\mbox{if}& -{\bf l}< x < 0 \\
V_{_{\infty}} &\mbox{if}& x >0,
\end{array}
\right.
\]
or constant, $V(x) = V_{\infty}, $ if ${\bf l}= 0$. The  first
case  corresponds to  presence of a  split-gate on the initial
part  $-{\bf l}< x < 0$ of the  wire, see below, section 3, in
second case the split-gate  is  absent or  switched  off.
\par
We  assume that the  tensor of effective mass in the  wires is
non-trivial: the effective mass across the wire $m^{^{\bot}}$ and
along the wire $m^{^{\parallel}}$ are  different, see the data for
Si above, or equal, as  for  some narrow gap semiconductors. The
magnitude of $m^{\bot}$ defines, up to some shift, the thresholds
$\frac{\hbar^2}{2m^{\bot}} \,\,\frac{\pi^2 l^2}{\delta^2},\,\,
l=1,2,\dots$, separating spectral bands in the wires.
\par
We  assume that the role of the one-electron Hamiltonian on the
network is played by the  Schr\"{o}dinger operator:
\begin{equation}
\label{Schredinger}
 l = - \frac{\hbar^2}{2 m}\bigtriangleup  +  V(x)
\end{equation}
with potential and tensor of effective mass  specified as
described above. It is  convenient  to  use  the  ``geometric"
form  of the corresponding  Schr\"{o}dinger equation  with
re-normalized spectral parameter $\lambda = p^{^2} =
2\,\frac{m_{_0}}{\hbar^{^2}}\left[ E - V_{_{\infty}} -
\frac{\hbar^{^2}}{2 m^{^{\bot}}}\frac{\pi^{^{2}}}{\delta^{^{2}}}
\right]$. On the  well  we  have:
\[
 - \frac{1}{m_{_0}}\,\bigtriangleup_x u(x)
+\frac{2}{\hbar^2}\,\left[V(x) - V_{_{\infty}} -
\frac{\hbar^{^2}}{2
m^{^{\bot}}}\frac{\pi^{^{2}}}{\delta^{^{2}}}\right]u =
\]
\begin{equation}
\label{geomSch} \frac{2}{\hbar^2}\left[E- V_{_{\infty}} -
\frac{\hbar^{^2}}{2
m^{^{\bot}}}\frac{\pi^{^{2}}}{\delta^{^{2}}}\right] u (x),\,=
\frac{1}{m_{_0}} \lambda  u ,\,|x|<R,
\end{equation}
with a new spectral variable  $\lambda$ or $p = \sqrt{\lambda}$,
which plays a role of an effective wave-number.
 The non-dimensional form of the above
Schr\"{o}dinger equation  on the  well is obtained via  change of
the space variable $x \to \xi = (\xi_{_1},\, \xi_{_2}) =
\frac{x}{R}$:
\[
-\frac{1}{m_0}\bigtriangleup_{\xi} u(R \xi)+ \frac{2\,
R^2}{\hbar^2} \left[V(R\xi) -  V_{_{\infty}} -
\frac{\hbar^{^{2}}}{2 m^{^{\bot}}}\,\,
  \frac{\pi^{^{2}}}{\delta^{^{2}}}\right]u(R \xi)
 =
 \]
 \begin{equation}
 \label{ndimSch}
 \frac{2 \,R^2}{\hbar^2} \left[E - V_{_{\infty}} - \frac{\hbar^{^{2}}}{2 m^{^{\bot}}}\,\,
  \frac{\pi^{^{2}}}{\delta^{^{2}}}\right]u(R \xi) =
R^{^2}\frac{\lambda}{m_{_0}} u(R\xi): =R^{^2}
\frac{p^{^{2}}}{m_{_0}} u
  ,\,\, |\xi|< 1.
 \end{equation}
The corresponding change of variables on each wire : $x \to \xi =
\frac{x}{R}$ along the  wire and $y \to \eta = \frac{y}{R}$ across
the  wire, $0< y < \delta $,\, $x >-{\bf l}$ gives  the  equation
:
\[
-\frac{1}{ m^{^{\parallel}}}\,\frac{d^{^{2}} u}{d \xi^{^{2}}} -
\frac{1}{m^{^{\bot}}} \frac{d^{^{2}} u}{d \eta ^{^{2}}} +
\frac{2\,R^{^{2}}}{\hbar^{^2}}\left[V - V_{_{\infty}}- \frac{1}{2
m^{^{\bot}}}\,\,\frac{\pi^{^{2}}}{\delta^{^{2}}} \right]u =
\]
\begin{equation}
\label{ndwires} \frac{2\,R^{^{2}}}{\hbar^{^2}} \,\left[E -
V_{_{\infty}} - \frac{\hbar^{^{2}}}{2 m^{^{\bot}}}\,\,
  \frac{\pi^{^{2}}}{\delta^{^{2}}}\right] u := \frac{R^{^2}}{m_{_0}} p^{^2} u =
  \frac{R^{^2}}{m_{_0}}\,\, \lambda\,\, u
\end{equation}
 We  will  use
further both geometric and  non-dimensional forms of
Schr\"{o}dinger equations  in the  wires  and  on the  well
assuming that each time proper change  of  variables  is also done
in the function $u$. We  also consider corresponding
Schr\"{o}dinger operators, in particular the  Intermediate
operator,  with  specified  coefficients and  boundary conditions,
 see the  next section 3 and  Appendix, section 9.1.
\subsection{Geometrical and physical limitations}
{\bf Spectrum } Our  approach  to  calculation of  transmission
coefficients is based  on introduction of an  ``Intermediate
operator''  defined as  Schr\"{o}dinger operator with  {\it
partial}
 ``chopping-off'' boundary  conditions  on the  bottom sections  of
channels, see (\ref{cho_p},\ref{match_2}) below. These  conditions
define  the Intermediate operator as a  Schr\"{o}dinger operator
(\ref{Schredinger}) on the orthogonal complement of the
 open channels in the  wires. If the Fermi level
sits  on the  first spectral band, then the non-trivial part of
the intermediate operator on the orthogonal complement of the open
channels  has a branch of absolutely-continuous spectrum (varying
multiplicity)
 beginning from the second
threshold in the  wires, $\sigma^{^r}_{_a} = \left[\left.
V_{_{\infty}}\, + \,\frac{4 \pi \hbar^{^2}}{2
m^{^{\bot}}\delta^{^2}},\, \infty\right)\right.$ and, probably,
few eigenvalues  below the second  threshold. Those of them which
sit above the  first threshold become embedded eigenvalues or are
transformed to resonances of the Schr\"{o}dinger operator on the
whole network, when the ``chopping-off'' boundary conditions
(\ref{cho_p},\ref{match_2}) are replaced  by the  matching
boundary conditions in all channels, see (\ref{match1}) below.
Submitting the standard scattering Ansatz to these  boundary
conditions (\ref{match1}) we will derive in the next section an
explicit expression (\ref{scattmatr}) for the Scattering matrix in
terms of the Dirichlet-to-Neumann map
 (DN-map, see Appendix, 10.2)  of the Intermediate operator.
It is  shown there that  the  ``partial'' DN-map of the
Intermediate operator is connected  with the standard DN-map of
the Schr\"{o}dinger operator on the quantum well, see  the
corresponding  formula (\ref{DNr}) and  the theorem 10.4 in
subsection 10.3. In particular the ``re-normalized" eigen-values
$\lambda^{^r}$  (of the intermediate operator )  which are sitting
on the  first spectral band,
 between the  first and  second
thresholds, are obtained by minor shifts $\Delta_{_s}$ of the
eigenvalues $\lambda_{_s}$ of the Schr\"{odinger} operator on the
quantum well with zero boundary condition  on the  whole boundary:
 \[
\lambda_{_s}\longrightarrow \lambda^{^r}_{_s} = \lambda_{_s} +
\Delta_{_s}.
 \]
The deviations of the  eigenvalues of the non-dimensional
Intermediate operator from  the eigenvalues of the corresponding
Schr\"{o}dinger operator on the  well with  zero  boundary
conditions are  estimated  in  Appendix, subsection  10.4. For
the most  interesting case of the  circular  well  the
corresponding non-dimensional shift  $R^{^2}\, \Delta_{_0}$ of the
resonance eigenvalue  was  estimated  as  $-0.02$,
\cite{Olejnik}.The restrictions of the corresponding
eigenfunctions of the Intermediate operator on the quantum well
coincide, in the lowest order of the approximation procedure, with
the eigenfunctions of the corresponding Schr\"{o}dinger operator
on the  quantum well, see Appendix, subsection 10.3.

In the  remaining part of this  section we  will not  distinguish
the  spectral data  of the  Intermediate  operator below the
second threshold from the spectral data of the Schr\"{o}dinger
operator on the  well with zero boundary  conditions. We assume
temporarily in this  section that split gates are  absent, ${\bf l
} = 0 $.
 \vskip0.5cm
\noindent{\bf Leading terms of DN-map} The  numerator and
denominator  of  the  above  approximation (\ref{Sapprox}) for the
Scattering  matrix  contain  the expression  for (\ref{DNess}) and
the  exponent $i{\bf p}I$. If the  Fermi level $E_{_{F}}$, and
hence  the  energy of  electron, sits  in  the upper  part  of the
first  spectral band, then  the  leading terms  in both numerator
and  denominator of the  approximate expression (\ref{Sapprox})
presented  in terms  of geometrical variables are: the resonance
term
\[
\frac{P_{_+}\frac{\partial \varphi_{_0}}{\partial n}\rangle
\langle P_{_+}\frac{\partial \varphi_{_0}}{\partial n}}{\lambda -
\lambda_0}
\]
which corresponds to the  ``resonance  energy level''
 \[
E_{_{0}} = \frac{\hbar^{^2} \pi^{^2}}{2 m^{^{\bot}}\, \delta^{^2}}
+ V_{_{\infty}} + \frac{\hbar^{^2} \lambda_{_0}}{2 m_{_0}}\approx
E_{_{F}},
\]
closest to  the  Fermi level  $F_{_{F}}$, and, probably the
exponent $i{\bf p }(\lambda)I$ which, in simplest case, is
proportional to  the effective wave-number  ${\bf p }(\lambda) =
\sqrt{\frac{m^{\parallel}}{m_{_0}}}\,\,p =
\sqrt{\frac{m^{\parallel}}{m_{_0}}}\sqrt{\lambda}$, see Appendix,
subsection 9.1 :
\[
p^{^2}_{_{0}} = \frac{2\,m_{_0}}{\hbar^{^2}}\left[ E_{_{0}}
 -  V_{_{\infty}} - \frac{\hbar^{^2} \pi^{^2}}{2 m^{^{\bot}}\,
\delta^{^2}}\right].
\]
The  subordinate (non- resonance) terms may be  estimated  by the
contribution to  ${\cal D}{cal N}_{_{T}}$ from the closest to
$E_{_{0}}$ ``non-resonance" energy level $E_{_{1}} \neq E_{_{0}}$:
\begin{equation}
\label{fraction}
 \frac{P_{_+}\frac{\partial \varphi_{_1}}{\partial
n}\rangle \langle P_{_+}\frac{\partial \varphi_{_1}}{\partial
n}}{\lambda - \lambda_1},
\end{equation}
and  by  the  contribution from the  upper branch of the
continuous spectrum  which begins  from the second threshold
$\sqrt{\frac{4 \pi^{^2} m^{^{\parallel}}}{\delta^{^{2}}\,
m^{^{\bot}}} + \frac{2 m^{^{\parallel}}
V_{_{\infty}}}{\hbar^{^2}}}$. This contribution  is estimated  by
the product of the inverse ``geometric spacing"  at the resonance
level  $\rho(\lambda_{_0}) = \min_{_{s}}|\lambda_{_{0}} -
\lambda_{_{s}} |$ and  the  norm $||P_{_+}\frac{\partial
\varphi_{_1}}{\partial n}||^{^2} = C_{_{R}}$ of the
one-dimensional operator
 ${\cal P}_{_1} = P_{_+}\frac{\partial \varphi_{_1}}{\partial n}\rangle \langle
P_{_+}\frac{\partial \varphi_{_1}}{\partial n} $  which  stays in
the numerator  of the  fraction (\ref{fraction}). It can be
estimated as $C_{_{R}} \leq \hat{C} R^{^{-3}}$ with a
non-dimensional constant $\hat{C}$. The  constant $\hat{C}$ is the
norm of the corresponding one-dimensional operator for circular
quantum well radius 1 , and  in the example  below it does not
exceed  10.
\par
We say, that the width of the wires is {\it relatively small
compared with the radius $R$ of the  quantum well}, if  the
inverse  ``re-normalized"  spacing on Fermi level is dominated by
the non-dimensional effective wave-number $p^{^{1}} = Rp$ :
\begin{equation}
\label{protrus2}
 \frac{\hbar^2}{2m_0 R^2 \rho^{^r} (E_{_F})} <<
  R
\sqrt{\frac{2m^{^{\parallel}} \left[ E_{_F} -
V_{_{\infty}}\right]}{\hbar^2}-
\frac{m^{^{\parallel}}}{m^{^{\bot}}} \frac{\pi^2 }{\delta^{^2}}
}:=
 \sqrt{\gamma_{_1}\,\,\frac{m^{^{\parallel}}}{{m_{_0}}}\,\,p^{^{1}}}
 = \frac{\pi \, R}{\delta}\,\,\sqrt{\gamma_{_1}
3\,\frac{m^{\parallel}}{m^{\bot}}}.
\end{equation}
The  right  side  of  the  above inequality and  the
``re-normalized" spacing $\frac{ 2m_{_{0}} R^2}{\hbar^2} \rho^{^r}
:= \rho^1 $ in the denominator in the left  side of the last
condition are non-dimensional. The ``re-normalized" spacing is
actually equal to the spacing on the resonance level
$\hat{\lambda_{_0}} : \frac{\hbar^2}{2 m_0
R^2}R^{^{-2}}\hat{\lambda_0}  = E_0 \approx E_{_F}$ of the
corresponding non-dimensional Schr\"{o}dinger equation.
\vskip0.5cm \noindent {\bf Temperature} When assuming that {\it
the radius $R$ of the quantum well is relatively small, for given
temperature}, we actually have in mind that the spacing $\rho
(E_F)= \min_{s \neq 0}|E_s - E_0|$ of energy levels in the quantum
well $\Omega_0$ radius $R$ at the resonance energy level $E_0$
closest to the Fermi level, $E_{_{0}} \approx E_F$ (with wires
disjoint from the domain) is large compared with temperature:
\begin{equation}
\label{protrus1} \kappa T < \frac{1}{2}\mbox{inf}_{_{_{E_s \neq
E_F}}}\,|E_F - E_s|  = \rho (E_{_0}).
\end{equation}
Generally, there  may be  several eigenvalues $E_{_s} = V_{\infty}
\frac{\pi^{^2}{\hbar}^{^2}}{2 m_{_0}}\lambda_{_s}$ of the
 Intermediate operator (or  the  Schr\"{o}dinger operator
 on the  well) situated  on the  essential interval of energy
centered  at  the  Fermi level $[E_{_{F}} - \kappa T ,\, E_{_{F}}
+ \kappa T ]$
\begin{equation}
\label{ess_intd} E_{_{F}} - \kappa T \leq  E_{_s} \leq E_{_{F}} +
\kappa T.
\end{equation}
In this  case  we  neglect the polar  terms  in  the  DN-map which
correspond to eigenvalues  outside  the  essential interval.
\par
\par
When assuming  that the  Fermi level  $E_F$ is situated in  the
upper part of the first spectral band $\left[
\frac{\hbar^2}{2m^{\bot}} \frac{\pi^2}{\delta^2} + V_{\infty},\,\,
\frac{\hbar^2}{2m^{\bot}} \frac{4 \pi^2}{\delta^2} +
V_{\infty}\right]$ in the wires, we  suppose  that  $E_{_{F}}$
divides  the  first spectral band  in ratio $\mu_{_{1}}:
\mu_{_{2}},\,\,\mu_{_{1}} + \mu_{_{2}} = 1,\, 0< \mu_{_2} <
\mu_{_1}$
\begin{equation}
\label{middle} E_F = V_{_{\infty}} + \mu_{_1}\frac{\hbar^2\,
\pi^{^2}}{2 m^{^{\bot}}\delta^{^2}} +  \mu_{_2}\frac{4\pi^2\,
\hbar^2}{2 m^{^{\bot}}\delta^2}.
\end{equation}
 Then one  can estimate
the distances of the  Fermi level
 to  the  first $\displaystyle
\frac{\hbar^2 \pi^2}{2 m^{\bot} \delta^2} + V_{\infty} $
 and the second $\displaystyle \frac{4\hbar^2 \pi^2}{2 m^{\bot}
\delta^2} + V_{\infty}$ thresholds  in the wires respectively as $
\mu_{_{1,2}}\,\, 3\,\,\frac{\hbar^2 \pi^2}{2 m^{\bot} \delta^2} $
and the square ``effective  wave-number'' on  on Fermi level as $
\displaystyle
 p_{_{F}}^{^2} =  3
\mu_{_{1}}\,\frac{m{_{_0}}\,\pi^2}{m^{\bot} \delta^2}$.
\par
Extending  the  principle  (\ref{ess_intd}) to  the  continuous
spectrum, we  will neglect the  contribution to  DN-map from the
the  continuous spectrum  of the  Intermediate operator if it
does-not overlap with  essential  spectral interval  for  given
temperature  $T$. If  the  Fermi level sits  below  the  second
threshold, this means :
\begin{equation}
\label{n_overlap} \gamma_{_2}\,\frac{3 \pi^{^2} \hbar^{^2}}{2
m^{^{\bot}}\, \delta^{^2}} > \kappa T.
\end{equation}
For  Si  with $m^{^{\bot}} = 0.19$  and  wires  width  $5$ nm this
condition estimates the  temperature in  K  as $ T < 250 K
\mu_{_2}$, which  means that  Nitrogen temperature  $77\,\, K$ are
low  enough to  neglect in approximate  formula  for  the
Scattering matrix  the  contribution from  the  second  spectral
band, if the  Fermi level divides  the  first  spectral band  in
ratio  $\mu_{_1} : \mu_{_2} =  2 : 1$. For  wires  width  $6 nm$
one  can  neglect  the contribution from  the continuous spectrum
of the  Intermediate operator  for Nitrogen temperature if the
Fermi level is  situated  in the  middle of the  first spectral
band.  Note  that the  width  of the  wires  as self-assembled
patterns  on the  Si matrix  can be already controlled  with
precision better  than  2 nm, \cite{B1} \vskip0.5cm
\subsection{Example}
In anticipation  of the discussion of  parameter's regime of the
Resonance Quantum Switch in section 6, we consider the most
interesting example of the switch, based on a circular quantum
well with  quantum wires  width  $\delta = R/2$.
 Assume  that  the  shift potential
$V_{_{0}}$ in the well is  selected  such that $V_{_{0}} -
V_{_{\infty}} - \frac{\hbar^{^2}}{2 m^{^{\bot}} \delta^{^{2}}} = 0
$. Then the  potential of the  corresponding non-dimensional
Schr\"{o}dinger equation on the  well is  just proportional
$\langle \xi,\, \nu \rangle u$. Assume that the re-normalized
electric field is selected as ${\bf e}= 18.86$. Then the
eigenfunction of the non-dimensional Schr\"{o}dinger operator on
the quantum well
\[
- \bigtriangleup_{_{\xi}} u  - {\bf e} \langle \xi,\, \nu \rangle
u = R^{^2}\,p^{^2}\, u,
\]
 which corresponds to the (non-dimensional)
resonance  eigenvalue $R^{^2}\lambda_0 = 14.62$
 has a special shape which  can  be used
for manipulation  of the quantum current, see  \cite{MP01} and the
section 5 below.
\par
We  assume  that $R^{^2}\lambda_0 = 14.62 = \hat{\lambda_{_0}}$ is
the non-dimensional resonance level, $\frac{\hbar^{^2}
\lambda_{_0}}{2 m_{_0}} \approx E_{_{F}}$ in the  well with zero
boundary  conditions. The  non-dimensional deviation $\Delta_0 = -
0.07$ from the corresponding ``shifted''  eigenvalue $
R^{^2}\lambda_0^r = 14.554$ of the intermediate operator  is small
and is dominated by the non-perturbed spacing - the  distance
 $- 2.30$ to the  nearest  non-resonance eigenvalue $12.32$. The above
condition (\ref{protrus2}) remains valid  for  the Intermediate
operator too. \vskip0.5cm \noindent {\bf Small parameter} If the
Fermi-level divides the  first spectral band in ratio
$\gamma_{_{1}}:\gamma_{_{2}} $, then  the one-pole approximation
is applicable to the switch with ``relatively narrow"  wires  if
the  condition  (\ref{protrus2}) is  fulfilled. In actual case  of
the switch based on a  circular  quantum well  this  condition
takes the  form:
\begin{equation}
\label{relsmall} \frac{1}{2.3} << \frac{R}{\delta}\,\pi
\sqrt{3\frac{m^{\parallel}}{m^{\bot}}} \sqrt{ \mu_{_1}},
\end{equation}
It obviously holds if the width $\delta$ of the wires does not
exceed $\frac{R}{2}$, and  the  Fermi level sits in the middle of
 the  first spectral band  $\mu_{_1} = \mu_{_2} = 1/2$:
It may be  reduced  in this  case to :
\begin{equation}
\label{rsm}
 1 << 20\sqrt{\frac{m^{\parallel}}{m^{\bot}}},
\end{equation}
which  is not restrictive at all in case $\delta < R/2$. Moreover
it  permits to  use  the ``natural'' small parameter
$\frac{1}{20}\,\sqrt{\frac{m^{\bot}}{m^{\parallel}}} $ in
perturbation procedures in sections  3,4 and  sub-sections 10.2,
\, 10.3.
\par
When formulating the basic condition
(\ref{protrus2},\,\ref{relsmall})  we deliberately  omitted  the
constant $\hat{C}$ which estimates the  norm of the
one-dimensional  operator in the  numerator of  the non-resonance
polar term  of  the  ${\cal D}{\cal N}_{_{T}}$. In our  example
the non-dimensional constant $\hat{C}$ is less  than 10, see the
sub-section 5.2 and  \cite{Robert}, hence  the accurate
estimation  of the non-resonance
 contribution {\it with regards of
the  constant $\hat{C} = 10$} still results in ``small'' parameter
$(4.4)^{^{-1}}$:
\begin{equation}
\label{smallpar} 10 << 20
\sqrt{\frac{m^{^{\parallel}}}{m^{^{\bot}}}} \approx 44.
\end{equation}
for Si, if $\delta \leq \frac{R}{2}$, which  shows  the  ``degree
of domination'' of  the  non-resonance contribution by  the
effective wave-number. For more  sophisticate  geometry  the
question on the  domination  requires  accurate  calculation.
\vskip0.5cm \noindent {\bf Low temperature} Though the condition
(\ref{protrus2}) was initially formulated independently of
temperature, we  saw that physically this condition is  required
only for
 spectrum of the intermediate  operator  inside  the
essential  interval of energy: $( E_{_F}- \kappa T,\, E_{_F} +
\kappa T)$. In particular, if  the  condition  (\ref{n_overlap})
is  fulfilled , this  interval does not overlap with upper
branches of the absolutely-continuous spectrum:
\begin{equation}
\label{temp1} E_{_{F}} + \kappa T < V_{_{\infty}} +
\frac{\hbar^{^{2}}}{2 m^{^{\bot}}} \,
\frac{4\pi^{^{2}}}{\delta^{^2}}.
\end{equation}
Then  the  contribution to  the essential part
 DN-map  of the  Intermediate  operator,
 and  hence to the  Scattering matrix (\ref{scattmatr}), from the
 upper  branches of the  absolutely-continuous  spectrum can be  neglected.
If there  are  no  eigenvalues  in the  essential interval of
energy,
\begin{equation}
\label{temp2}
 E_{_F}- \kappa T < E_{_{0}} < E_{_F} + \kappa T,\,
\end{equation}
except the  resonance  eigenvalue $E_{_{0}}\approx E_{_{F}}$, that
is $|E_{_{F}} - E_{_{t}}|\geq \kappa T$ for  $t \neq 0$ ,\, then
only  one  resonance term  remains in the  corresponding
approximate
 expression  (\ref{DNess}) for  the  DN-map, and  hence
for  Scattering matrix the `` one-pole  approximation''
(\ref{onepole}) can be  used, see the  discussion in  section 2.
If all these  condition (\ref{protrus1},
\ref{protrus2},\ref{temp1},\ref{temp2}) are fulfilled  (on  the
essential  interval of energy), then the transmission coefficient
from one wire to another exhibits clear resonance properties in
single-mode scattering process see  the  next sections 4,5. These
properties are defined by the shape of the resonance
eigenfunction, see the discussion in section 5. In
 section 4, using the one-pole approximation for  the  DN-map
Scattering matrix, and  the presence of the  small  parameter
arising from  (\ref{smallpar}) we calculate the resonances  and
estimate  the speed  of switching.
 \vskip0.5cm
\section{Boundary conditions and
Intermediate operator.\\Scattering matrix.}

\subsection{Split-gates}
We consider a single act of the electron's transmission from the
incoming wire to one of terminals as a scattering process on a
quantum network constructed of a circular quantum well
 $\Omega_0 = \left\{x : |x|<R \right\}$ and several
straight quantum  wires - the trenches $\Omega_n$  width $\delta$
- attached to it orthogonally centered at the points $a_n$ on the
boundary of the well ( $n = 1,2,3$ for two-terminal switch RQS-2
and $n = 1,2,3,4$ for three-terminal switch RQS-3). The wire
$\Omega_1$ is selected for the input, others are terminals.
Positions of the points $a_n$ are  chosen to optimize the
switching effect, see below, section 5. One may suggest two
engineering solutions of the problem of manipulation of the
quantum current across the  switch. The first may  be based on
manipulation of the hight  of  barriers $V_{barrier} =
\frac{\hbar^{^2}H^{^2}}{2 m_{_o}}$ on the  initial part  of the
wires $-{\bf l}< x < 0$ via the change of the corresponding
electric field, thus closing or opening the ``split-gates" . This
mechanism  may be used even in case when the  wires  are wide
enough (not ``single  mode"). The speed of transition processes in
the switch ( hence the speed of switching) in this case  depends
not  only on  geometry of  the network, but is defined  by  the
microscopic properties of  the materials and can't  be  easily
estimated  theoretically. Another mechanism is based on the
resonance scattering in the switch. It can be applied only to
single-mode wires. The speed  of switching in  second case  can be
estimated in terms of geometrical parameters of the switch. Though
the first mechanism is efficient for manipulating stronger
currents, we consider in this paper the second one, based on
solution of the resonance scattering problem for the corresponding
Schr\"{o}dinger equation
\[
-\frac{\hbar^2}{2m_0}\bigtriangleup u + V(x)u = E u.
\]
\vskip0.5cm
\subsection{Optimization of design: solvable model versus scanning}
If all parameters of the model are chosen, then the above
scattering problem can be, at  least ``in principle'', solved
numerically with proper potential $V$ on the whole  space when
assuming  that the potentials and  tensor of  effective  mass on
the  network. Mathematical theory of the above scattering problem
on the network with zero boundary condition is developed in
\cite{MP02,P02,QSW}. In actual paper we use presence of the
natural small  parameter, which  permits, practically, to neglect
the contribution  to  the  Scattering matrix from the  non -
resonance
 eigenvalues. Moreover we may assume that the
matching  of the  solution of the Schr\"{o}dinger equation on the
circular quantum well with the solutions in the rectangular wires
may  be done on  the flat bottom sections just via  replacement of
the  central angle  by  the  corresponding  sine  due  to  the
 condition  $ \delta/R < 1/2,\, \theta < \pi/ 6$:
\[
\theta -\sin \theta \leq \frac{\pi}{6} - \frac{1}{2} \approx
0.023,
\]
This observation  permits to calculate  approximately the
resonance entrance  vectors  forming  projections in  the
numerator of polar terms  of  the  DN-map  (\ref{DNess}), see
subsection 10.3, and  even  estimate  accurately the error
appearing from substitution  of the circular quantum  well
$\Omega_0$  by the  corresponding ``almost circular'' modified
domain $ \Omega \backslash \left\{\Omega_1\cup\Omega_2\cup\dots
\right\}$ which has  flat pieces on the  boundary  matching the
bottom sections of the wires, see  \cite{P02}. Results  of the
spectral analysis of  the Schr\"{o}dinger  equation on the
modified quantum well with flattened pieces at  the  places  of
contact with wires, only slightly deviate from ones on the
circular one.
\par
 Nevertheless  neither numerical solution, nor the straightforward
experimental search  are efficient in course of search  of the
parameter regime  where the Resonance Quantum Switch  works, since
the space of physical and geometrical parameters of the switch is
multi-dimensional. The ``working point'' of the  switch  is
actually a sort of a ``multiple point" which is easy to miss when
scanning on one of  parameters with other parameters  fixed at
random. Our search of the working point is based on mentioned
above explicit approximate formula  (``one-pole approximation")
for  the  Scattering  matrix which permits to obtain preliminary
estimation of  the working parameters, thus minimizing  the  task
of  the  random  search. The  derived approximation is  actually
an exact Scattering matrix of proper solvable model which is
constructed in section 7. \vskip0.5cm
\subsection{Intermediate operator}
In this sub-section we  announce
 an exact formula for the  Scattering matrix based  on
Dirichlet-to-Neumann map of the Intermediate  operator. The
corresponding elementary calculation is  postponed to  the
subsection  9.2. We derive  the  formula for the geometric form of
the  Schr\"{o}dinger equation on the network,  replacing the
standard  Schr\"{o}dinger equation (\ref{Schredeq}) on the well
$\Omega_{_{0}}$ by
\begin{equation}
\label{Schred_G}
  -\frac{1}{m_{_{0}}}\bigtriangleup u + \frac{1}{m_{_{0}}}{\bf V}_{_{0}}(x)u =
  \frac{1}{m_{_{0}}} \lambda u,
\end{equation}
and  on  the  wires  $\Omega_{_{s}}$ by
\begin{equation}
\label{Schred_GW}
  -\frac{1}{m^{^{\parallel}}}\bigtriangleup u + \frac{1}{m^{^{\parallel}}}{\bf V}_{_{s}} (x)u =
  \frac{1}{m_{_{0}}}\lambda u,
\end{equation}
see  (\ref{ndimSch},\ref{ndwires}) and Appendix, where all
relevant  notations are  introduced, (\ref{welleq},\ref{wireq}).
\par
If the Fermi energy sits on the  first spectral band
 it  is convenient to use an {\it
Intermediate} perturbed  operator $l^{^r}_{_0}$ defined as
Schr\"{o}dinger operator  on  the  whole network with  proper
matching  conditions  in  all  channels except  the first one,
where the matching boundary  condition (\ref{match}) is replaced
by the
 {\it partial} Dirichlet  boundary condition,
{\it chopping  the  first  channel off}. For  the  three-terminal
Resonance Quantum Switch we  denote by  $ E_{_{+}}$ the
$4$-dimensional entrance subspace of open channel, spanned  by
the first-order eigenfunctions
 $e^{^s}_{_1},\,s=1,2,3,4,$ on the
bottom sections ${\bf\gamma_s}\, s=1,2,\dots,\,$ of the wires of
the wires, $e_{_{1}} = \sqrt{\frac{2}{\delta}}\, \sin\frac{\pi
y}{\delta}$, and by $P_+$ the corresponding orthogonal projection
in $L_2(\Gamma)$. Then  denoting  the  sum  of  the  bottom
sections $\gamma_{_s}$ by  $\cup {\bf\gamma_{_{s}}} = \Gamma $ we
present the partial {\it chopping-off} boundary condition  as
\begin{equation}
\label{cho_p} P_+  u\bigg|_{_{\Gamma}} = 0,
\end{equation}
both for the functions from the  domain of the corresponding split
operator  in the  wire and  in the  well. The partial matching
condition in all upper ( closed) channels with the entrance
subspace  $E_{_-}$ and  the corresponding complementary projection
$P_- = I_{\Gamma} \ominus P_+$ in $L_2(\Gamma)$ is taken in form :
\begin{equation}
\label{match_2} P_- \left[u - u_0\right]\bigg|_{\Gamma}= 0,\,\,
P_-\left[\frac{1}{m^{\parallel}}\frac{\partial u}{\partial n} -
\frac{1}{m_0} \frac{\partial u_0}{\partial
n}\right]\bigg|_{\Gamma}= 0.
 \end{equation}
The  split operator  $l^{^{r}}$ defined  by the  above
 differential expressions  (\ref{Schred_G}, \ref{Schred_GW}) and  the
 boundary  conditions  (\ref{cho_p},\ref{match_2}) can be  presented  as
an orthogonal sum  of the {\it trivial part}: the one-dimensional
Schr\"{o}dinger operators  $l_{_{s}},\,\, s=1,2,3,4$ :
\[
l_{_{s}} = -\frac{1}{m^{^{\parallel}}}\,\,\frac{d^{^2} u_{_{s}}}{d
x^{^{2}}} =
 \frac{1}{m_{_{0}}} \lambda u_{_{s}},
\]
on the open channels with zero boundary  conditions  at the bottom
sections, and  the {\it non-trivial  part}  $l_{_{0}}^{^{r}}$
defined in the  orthogonal complement  in Hilbert space of all
square-integrable  functions  on the  network $l^{^{r}} =
\sum_{_{s = 1}}^{^{4}} l^{^r}_{_{s}}\oplus l^{^r}_{_{0}} $.
\par
We  will  use  for  the  Intermediate  operator the  geometrical
rescription and  the geometrical spectral
 parameter  $\lambda = p^{^{2}}$ introduced  in the  previous  section. Then
the continuous  spectrum  of  the component $l^{^r}_{_{0}}$  of
the intermediate  operator  on the orthogonal complement of the
blocked  first  channel  begins from the  second threshold
$\frac{m_{_0}}{m^{^{\bot}}} \frac{3 \pi^2 }{\delta^2}  $.  Later
we will  reveal  a  special role played  by the the eigenvalues
$\lambda^r_s$ of $l^{^r}_{_{0}}$ sitting on the  first spectral
band  $0< \lambda < \frac{3\pi^{^2}}{\delta^{^2}}$.
 \par
The intermediate scattering  problem may be convert into the
original scattering problem via replacement of the partial zero
condition (\ref{cho_p}) in the  first  channel by  the
corresponding partial matching condition:
\begin{equation}
\label{match1} P_+ \left[u - u_0\right]\bigg|_{\Gamma_s}= 0,\,\,
P_+\left[\frac{1}{m^{\parallel}}\frac{\partial u}{\partial n} -
\frac{1}{m_0} \frac{\partial u_0}{\partial
n}\right]\bigg|_{\Gamma_s}= 0.
 \end{equation}
The  perturbation caused  by  this  one-dimensional change  of the
boundary  condition (\ref{cho_p}) to (\ref{match1}) transforms the
separated branch of  continuous spectrum  in  the  split first
channel $0< \lambda <\infty$ into the  branch of continuous
spectrum of  the  original spectral problem. The corresponding
scattered  waves are  combined of Jost solutions $f_{\pm} = e^{\pm
i\sqrt{\frac{m^{\parallel}}{m_0}}\,p\, x}, x>0, $ of  the equation
(\ref{separ}) with compactly-supported potential $\left[ V_1 (x) -
V_{\infty}\right] = 0 ,\,\, x>0$ and $\left[ V_1 (x) -
V_{\infty}\right]= H^{^2} \frac{\hbar^2}{2m_0},\,\, -{\bf l}<x<0$:
\[
-\frac{d^2 u_1}{d x^2} + \frac{2 m^{\parallel}}{\hbar^2}
\left(V_1(x) - V_{\infty} \right)
 u_1 = \frac{m^{\parallel}}{m_0}p^2 u_1,
\]
\begin{equation}
\label{ansatz} u_1(x)= e^{-i\sqrt{\frac{m^{\parallel}}{m_0}}px} e
+
 e^{i\sqrt{\frac{m^{\parallel}}{m_0}}px} S_1 e ,
\end{equation}
for $l=1,\, x>0,$ (on the  open channel), with any vector $e \in
E_{_{+}}$, accomplished with  exponentially decreasing  components
in  upper channels $l> 1$:
\[
u_l (x) =  e^{-\sqrt{\frac{m^{\parallel}}{m_0}}
\sqrt{\left(\frac{\pi^2 l^2}{\delta^2} - \frac{\pi^2 }{\delta^2}
\right)- p^2}\,\,x} S_l e,\,\,x>0,\,\,l > 1.
\]
Here $S_1$- the  Scattering Matrix - and the amplitudes
$S_l,\,l>1,$ in  upper  channels are  defined from the  matching
condition  (\ref{match1}) of $u_1$ to the  solutions of the
corresponding homogeneous equation inside the  well. \vskip0.5cm
\subsection{Scattering matrix}
In standard techniques  one  usually  matches the  solution  of
the Schr\"{o}dinger equation in the domain with  the  solutions in
all (open and  closed) channels  in the  wires. We  suggest, see
Appendix, sub-section 10.2, a partial matching  technique, which
requires  matching  of  solutions of the homogeneous intermediate
equations with  the solutions {\it in open channels of the  wires
only}, thus  eliminating infinite algebraic system. Mathematical
convenience  of  this approach, see \cite{QSW} and  Appendix
below,
 consists  in  eliminating of  unbounded  operators,
 which  are  replaced in our approach by  finite  matrices.
Using the partial  matching approach  based  on  the intermediate
operator gives  the  following  explicit formula for the
scattering  matrix in case  when  the  split-gate is  present:
 \begin{equation}
 \displaystyle
\label{scattmatr}
 S (p)= - \frac{i\sqrt{\frac{m^{\parallel}}{m_0}}p
 \frac{\tanh\sqrt{\frac{m^{\parallel}}{m_0}(H^2 -
  p^2)}\,\,\,{\bf l}}{\sqrt{H^2 -
 \frac{m^{\parallel}}{m_0} p^2}}\,+\,1}
 {-i\sqrt{\frac{m^{\parallel}}{m_0}}p
 \frac{\tanh\sqrt{\frac{m^{\parallel}}{m_0}(H^2 -
  p^2)}\,\,\,{\bf l}}{\sqrt{\frac{m^{\parallel}}{m_0}(H^2 -
  p^2)}\,+\,1}}\,\,\,
 \frac{\frac{m^{\parallel}}{m_0}\,P_+ \Lambda^{^r} P_+ - P_+{\bf Q}_{_{\bf l}}}
 {\frac{m^{\parallel}}{m_0}\,P_+ \Lambda^{^r} P_+ - P_+\overline{\bf Q}_{_{\bf l}}},
\end{equation}
where
\[
{\bf Q}_{_{\bf l}} = \frac{-i\sqrt{\frac{m^{\parallel}}{m_0}}p -
\sqrt{\frac{m^{\parallel}}{m_0}(H^2 -
  p^2)}
 \tanh\sqrt{\frac{m^{\parallel}}{m_0}(H^2 -
  p^2)}\,\,\,{\bf l}}
 {i\sqrt{\frac{m^{\parallel}}{m_0}}p
 \frac{\tanh\sqrt{\frac{m^{\parallel}}{m_0}(H^2 -
  p^2)}\,\,\, {\bf l}}{\sqrt{\frac{m^{\parallel}}{m_0}(H^2 -
  p^2)}}\,+\,1}
\]
and $\Lambda^r$ is  the  Dirichlet-to-Neumann  map (DN-map), see
\cite{SU2,DN01}, of  the intermediate  operator $l^r$, see
Appendix. The  corresponding  formula  for  the case  when the
split-gate is  absent can be  obtained  via  replacement the width
${\bf l}$  of the  barrier by zero, ${\bf Q}_{_{0}} = - i
\sqrt{\frac{m^{\parallel}}{m_0}} p $.
\par
The  resonance  properties  of the  Scattering  matrix
(\ref{scattmatr})  may be  revealed  when  substituting  the
spectral series/integral
 for  the  $\Lambda^{^r} = P_+ \Lambda^r P_+$
\[
\Lambda^{^r} = \sum_{_{\lambda_s}} \frac{P_+
\frac{\partial\varphi^{^r}_{_s}}{\partial n}\rangle \langle P_+
\frac{\partial\varphi^{^r}_{_s}}{\partial n}}{\lambda -
\lambda^r_s} + \int_{\sigma > \frac{4 \pi^2}{\delta^2}} \frac{P_+
\frac{\partial\varphi^{^r}_{_2} (\sigma)}{\partial n}\rangle
\langle P_+ \frac{\partial\varphi^{^r}_{_2}(\sigma)}{\partial n}}
{\lambda - \sigma} d \sigma + \int_{\sigma > \frac{9
\pi^2}{\delta^2}} \frac{P_+ \frac{\partial\varphi^{^r}_{_3}
(\sigma)}{\partial n}\rangle \langle P_+
\frac{\partial\varphi^{^r}_{_3}(\sigma)}{\partial n}}{\lambda -
\sigma} d \sigma + \dots.
\]
Practically the  calculation of the DN-map and  the Scattering
matrix for low  temperatures can be reduced to  the  calculation
of the  contribution from the singularities of the  DN-map of the
intermediate  operator  sitting on the
 the essential interval of energy, see (\ref{temp2}).
 In the  next  section  we neglect the  contribution from
 the  upper  branch  of the  continuous  spectrum  of the  intermediate
 operator  and  estimate  the  contribution to
 the  DN-map  (and  to  the  Scattering matrix) from the non-resonance  terms, $\lambda_s
\neq \lambda_0$  by the  inverse spacing $\rho^r(\lambda^r_0)$ on
the  resonance level $\lambda = \lambda^r_0$ and some  dimensional
constant  $ C \approx [\mbox{length}]^{^{-3}} $ introduced in
section 2.2:
\begin{equation}
\label{specDN}
\parallel \frac{P_+ \frac{\partial\varphi^{^r}_{_s}}{\partial n}\rangle
\langle P_+ \frac{\partial\varphi^{^r}_{_s}}{\partial n}}{\lambda
- \lambda^r_s}\parallel =
O\left(\frac{1}{\rho^{^{r}}}(\lambda^{^r}_{_0})\right) \leq
\frac{C}{|\lambda^r_0 - \lambda_s|}\leq \frac{C}{\rho^r
(\lambda^r_0)}.
\end{equation}
In the  next  section we will show  that  the  most  interesting
case  corresponds to  ${\bf l} = 0$. Then the  numerator  of the
scattering matrix  is  presented for  low  temperatures as:
\begin{equation}
\label{numerator} \frac{P_+
\frac{\partial\varphi^{^r}_{_0}}{\partial n}\rangle \langle P_+
\frac{\partial\varphi^{^r}_{_0}}{\partial n}}{\lambda -
\lambda^r_{_0}} + O\left(\frac{1}{\rho^{^{r}}
(\lambda^{^r}_{_0})}\right) + i
P_{_{+}}\,\,\sqrt{\frac{m^{^\parallel}}{m_{_0}}}\,\, p.
\end{equation}
Leading terms  in the  numerator  near the  resonance  eigenvalue
$\lambda^r_{_0}$ of the  intermediate operator are  the  polar
term  $\frac{P_+ \frac{\partial\varphi^{^r}_{_0}}{\partial
n}\rangle \langle P_+ \frac{\partial\varphi^{^r}_{0}}{\partial
n}}{\lambda - \lambda^r_{_0}}$ and the  last term $i
P_{_{+}}\,\,\sqrt{\frac{m^{^\parallel}}{m_{_0}}}\,\,\, p$
containing the effective  wave number  $p$. Both of them are
homogeneous  functions  degree  $-1$ of the
 space variable. The middle term defining  the  contribution
 $O\left(\frac{1}{\rho^{^{r}}(\lambda^{^r}_{_0})}\right)$\
from the  non-resonance eigenvalues  is  also  a  homogeneous
operator-function degree $-1$ and can be  neglected  if  the
condition  (\ref{protrus2}) is fulfilled. In fact  one  can also
 develop  the  perturbation technique for  calculation  zeroes
 of (the  numerator)  of the  Scattering matrix based  on
 the  small parameter  discovered  in  sub-section 2.3.
\par
 If the  conditions  (\ref{temp1},\ref{temp2}) are  fulfilled, then  the contribution
to  the  DN-map from the non-resonance terms and   upper  branches
of the absolutely  continuous  spectrum  can be  neglected
resulting  in  the  resonance
 term only:
 \[
\Lambda^{^{r}} (\lambda) \approx  \frac{P_+
\frac{\partial\varphi^{^r}_{_0}}{\partial n}\rangle \langle P_+
\frac{\partial\varphi^{^r}_{_0}}{\partial n}}{\lambda -
\lambda^{^r}_{_0}}.
\]
This gives a convenient ``one-pole" approximation  for  the
essential DN-map, ${\cal D}{\cal N}_{_{T}}$, and the corresponding
``one-pole approximation" for the Scattering matrix which  is
used in the  next  section, see (\ref{onepole}) below. In the
next section we  use  also the above estimation (\ref{specDN}) for
neglected non-resonance terms. \vskip1cm
\section{The Life-time of  resonances  and  the speed of  switching}
\subsection{Life-time}
The $4\times4$ Scattering  matrix (\ref{scattmatr}) in the first
channel is an analytic matrix-function in complex  plane of
effective wave-number $p$ and may  have zeroes
 - resonances  $\left\{p_s\right\}$- in the upper  half-plane  and complex-conjugate
poles in the lower half-plane of  the complex  plane  of  $p$. In
particular,
 when the  split-gate is  absent, the  resonances can be  found  as  vector-zeroes
 of the  numerator (\ref{numerator})
\begin{equation}
\label{vectzeroes}\frac{P_+
\frac{\partial\varphi^{^r}_{_0}}{\partial n}\rangle \langle P_+
\frac{\partial\varphi^{^r}_{_0}}{\partial n}\, e\rangle}{\lambda -
\lambda^r_{_0}} +
 O\left(\frac{1}{\rho^{^{r}}
(\lambda^{^r}_{_0})}\right) e + i
\,\,\sqrt{\frac{m^{^\parallel}}{m_{_0}}}\,\,p \,\,e = 0 ,
\end{equation}
with a normalized vector $\,\, e \in E_{_{+}}$. Multiplying by the
orthogonal projection $P_{_{0}} =
 \frac{\phi_{_0}\rangle\,\,\langle
\phi_{_0}}{|\phi_{_0}|^{^2}}$ onto the  ``resonance entrance
vector" $ \phi^{^r}_{_0}: = P_+
\frac{\partial\varphi^{^r}_{_0}}{\partial n}: = \phi_{_0} =
\left\{\phi_{_0}^{^1},\phi_{_0}^{^2},\phi_{_0}^{^3},\phi_{_0}^{^4}
\right\} $ we  may reduce the equation to the  pair of equations
\[
\frac{|\phi_{_0}|^{^2}}{\lambda - \lambda^r_{_0}} +\langle
 P_{_0}\, O\left(\frac{1}{\rho^{^{r}} (\lambda^{^r}_{_0})}\right) e \rangle +
i \,\,\sqrt{\frac{m^{^\parallel}}{m_{_0}}}\,\,p  = 0,
\]
\[
-i \,\,\sqrt{\frac{m^{^\parallel}}{m_{_0}}}\,\,p\, (P_{_{+}} -
P_{_0})\,\,e + (P_{_{+}} =  P_{_0}) O\left(\frac{1}{\rho^{^{r}}
(\lambda^{^r}_{_0})}\right)\,e.
\]
The  first  equation is used to  estimate the  position of the
resonance $p $, see the  sub-section 2  below, and the  second can
be  used  to estimate   the  deviation  of the  corresponding
zero-vector
 $e$ from  the direction  of the ``resonance entrance vector''
 $ e_{_{0}} = |\phi_{_0}|^{^{-1}} \phi_{_0}$.
For given  resonance $p$ and  the corresponding  null-vector $e$ a
resonance  solution $u_{_0}$  of the  Schr\"{o}dinger equation
exists, with exponential asymptotic in the  wires:
\[
u_{_0} (x) = \left\{e^{^1},e^{^2},e^{^3},e^{^4} \right\} e^{^{-i
\sqrt{\frac{m_{_0}}{m^{^{\parallel}}}} \,\,\,p\,\,\, x}}
\]
The  corresponding  solution of the non-stationary Schr\"{o}dinger
equation
\[
\frac{\hbar}{i}\frac{\partial u}{\partial t} + \frac{\hbar^{^2}}{2
m_{_0}} \bigtriangleup u - V(x) u
\]
\begin{equation}
\label{lifetime} u(x,t) = e^{^{i\frac{\hbar}{2 m_{_0}} p^{^2} t}}
u_{_0}(x) :=
 e^{^{-\frac{t}{\tau}}} \,\, e ^{^{i\frac{\hbar}{2 m_{_0}} \Re{p}^{^2} t}} \,u_{_0}(x)
\end{equation}
is  exponentially  decreasing  with  the  decrement
 $\frac{1}{\tau} = \frac{\hbar}{2 m_{_0}} \Im{p}^{^2}$.
The  inverse decrement  $\tau$ is  called  the  life-time of the
resonance. The  life-time  is
 defined  similarly
for  split-gate  closed, as  zeroes of the  numerator of the
expression  (\ref{scattmatr}).
\par
Note  that  for dynamics associated with wave equation, see
\cite{Lax}, the life-time is  usually measured  by the inverse
imaginary  part of  the  resonance  in  the  plane  $p$ of the
wave-numbers.
 \par
In this section we will calculate the life-time of the  resonance
approximately estimating the  errors  appearing  from neglecting
of the  non-resonance terms, in  two  cases:
 \vskip0.5cm
 \noindent
1. In case when equivalent split-gates are constructed  on  the
initial part of each  wire. \vskip0.5cm
 \noindent
2.In case when  split-gates  are  absent (or  switched off) and
the  wires  are  attached straight  to  the  quantum well.
\subsection{Split-gates closed} Assume that the barrier formed by the split-gate is $1
ev$ over the Fermi level $E_f$ and  the  Fermi-level is  $1 ev$
over the effective bottom $V_{\infty} +
\frac{\hbar^2}{2m^{\bot}}$, and the width $\delta$ of the wire and
the width $l$ of the barrier both are $2\,\,\, nm$.Then, due  to
\begin{equation}
\label{value} \frac{2 m_0 \left(V_1 - E_f\right)}{\hbar^2}=
\frac{1}{3.81} \,\, \frac{1}{A^2},
\end{equation}
the ``effective momentum'' in the  wire , see  (\ref{separ}) and
below, is  estimated as
\[
p = \sqrt{\frac{2 m_0 \left(E_f - V_{\infty}\right)}{\hbar^2} -
\frac{m_0}{m^{\bot}}\, \frac{\pi^2}{\delta^2}} = 0.372
\frac{1}{A},
\]
and  the  under-barrier decay rate is  defined  by  the decrement:
\[
\sqrt{\frac{m^{\parallel}}{m_0}(H^2 - p^2)}\,=\,\, 0.45 \,\,\,
\frac{1}{A}.
\]
Then $ \sqrt{\frac{m^{\parallel}}{m_0}(H^2 - p^2)}\,\,{\bf l} = 9
$ and $\tanh \sqrt{\frac{m^{\parallel}}{m_0}(H^2 - p^2)}\,\,{\bf
l} = 1 - 2 e^{-18} = 1 - 3.45 \,\,10^{-8} \, := 1- \varepsilon$.
The approximate value of the fraction in the numerator may be
 estimated now  as
 \[
 \frac{m_0}{m^{\parallel}}
\frac{i\sqrt{\frac{m^{\parallel}}{m_0}}p \,+\,
\sqrt{\frac{m^{\parallel}}{m_0}(H^2 - p^2)}
 \tanh\sqrt{\frac{m^{\parallel}}{m_0}(H^2 - p^2)}\,\,{\bf l}}
 {i\sqrt{\frac{m^{\parallel}}{m_0}}p
 \frac{\tanh\sqrt{\frac{m^{\parallel}}{m_0}(H^2 - p^2)}\,\,{\bf l}}
 {\sqrt{\frac{m^{\parallel}}{m_0}(H^2 - p^2)}}\,+\,1}\,\,\,=
 \]
 \[
 = \frac{10}{9}\,\,\, \frac{i \,\,0.95 \,\times\, 0.372  +
  0.45(1 - \varepsilon)}
 {i \, 0.95 (1-\varepsilon)\,\, 0.372  +  0.45}\,\, 0.45
 \approx
\frac{1}{2}\,\, \left[ I - 0.246 \varepsilon + i\, 0.96
\,\varepsilon \right]
\]
 If the distance from the  resonance eigenvalue
 $E_{_{0}} \approx E_{_F}$  to the closest upper (second)
 threshold
 $ \frac{4 \pi^{^2}\hbar^{^2}}{2 m^{^{\bot}}\delta^{^2}}$ is  greater than $\kappa T$,
 then the contribution of  the  continuous spectrum to  the $DN$-map of the  intermediate
 operator may be  neglected  and   the zero of the  scattering  matrix closest  to
 resonance eigenvalue  can be  found  from the  equation
\begin{equation}
 \label{resonan}
\left[\frac{P_+ \frac{\partial\varphi^r_0}{\partial n}\rangle
\langle P_+ \frac{\partial\varphi^r_0}{\partial n}}{\lambda -
\lambda^r_0}\right] e  + \sum_{s\neq 0} \frac{P_+
\frac{\partial\varphi^r_s}{\partial n}\rangle \langle P_+
\frac{\partial\varphi^r_s}{\partial n}}{\lambda - \lambda^r_s}\, e
\,-
 \frac{m_0}{m^{\parallel}} {\bf Q}_{_{\bf l}} e = 0.
 \end{equation}
Here  $\varphi^r_s$ are the  normalized  eigenfunctions of the
Intermediate operator $l^r$ which  correspond  to  the
eigenvalues on the  essential  interval of energy. We pass  to the
non-dimensional  coordinates $\xi = R^{-1}x$,assuming that the
radius of  the  well $\Omega$ is  $250 A$. We will  use also the
non-dimensional spectral parameter: $\hat{\lambda} =
\lambda\,R^{2}$ and  spacing $\hat{\rho} = \rho \,R^{2}$. Then
based  on the crude estimate of the contribution to DN-map from
the  neighboring (non-resonance) eigenvalues by the input  from
the closest  neighbor :
\[
\parallel \sum_{s\neq 0} \frac{P_+ \frac{\partial\varphi^r_s}{\partial n}\rangle
\langle P_+ \frac{\partial\varphi^r_s}{\partial n}}{\lambda -
\lambda^r_s} \parallel \leq  R^{-1}
 O\left( \frac{1}{\hat{\rho}(\hat{\lambda^r_0})}\right)
\]
we may estimate the one-pole  approximation for the
non-dimensional DN-map  $\hat{\Lambda}^r$ near to  the
(non-dimensional) resonance:
\[
\Lambda^r = R^{-1} \hat{\Lambda}^{1,r} = R^{-1}
\frac{\hat{\phi}_0\rangle\,
 \langle \hat{\phi}_0}{\hat{\lambda} - \hat{\lambda^r_0}} + R^{-1}
 O\left( \frac{1}{\hat{\rho}(\hat{\lambda^r_0})}\right)=
\]
\[
R^{-1}\left[\frac{\hat{\phi}_0\rangle\,
 \langle \hat{\phi}_0}{\hat{\lambda} - \hat{\lambda^r_0}} +
  O\left( \frac{1}{\hat{\rho}(\hat{\lambda^r_0})}\right)\right].
\]
For eigenfunctions corresponding to the  neighboring eigenvalues
we  have  $|\hat{\phi}_s|^2 \approx 10,\,\,
\hat{\rho}\left(\hat{\lambda^r_0}\right) \approx 2.3$, hence the
contribution from the corresponding terms to  the  DN-map may be
estimated based  on  $\hat{C} <\ leq 10$ approximately as
 $ R^{-1}\frac{10}{2.3}= 4.3/ R $. We
will  use  for  this  quantity  the  notation  $R^{-1}
O\left(4.3\right) $. Then the equation for zeroes or the
Scattering matrix (in non-dimensional scale) may be presented  due
to  $R=250 A$ as:
 \[
0 = \frac{|\hat{\phi}|^2}{\hat{\lambda} - \hat{\lambda^r_0}} +
O\left(4.3\right)+ \frac{250}{2} \left[ 0.764  + i 0.96 \epsilon
\right] =
 \frac{|\hat{\phi}|^2}{\hat{\lambda} - \hat{\lambda^r_0}} + 95.5 +
O\left(4.3\right) + i \, 120\, \varepsilon.
 \]
Then, noticing that $\bigg|O\left(4.3\right)\bigg|<< 95.5$,
 {\it we may neglect the  contribution to DN-map  from the neighboring
 non-resonance eigenvalues} and obtain the (approximate) position of the
 zero of the  Scattering matrix
$\hat{\lambda}^r = \hat{p}^{^2}$ near to
 the  resonance eigenvalue   $\hat{\lambda}_0^r$ of the intermediate operator
just from the  one-pole approximation of DN-map:
\[
\hat{\lambda} \approx \hat{\lambda}_0^r - 0.11 + \, 0.14 i\,
\varepsilon \,\,
\]
Hence  due to $\varepsilon = 2\,e^{-18} = 10^{- 8}\,\, 3.45$ the
imaginary part of the  non-dimensional resonance is $10^{-8}\,\,
0.48$. The life-time may be calculated  now as
\[
\tau = \frac{2m_0}{\hbar}\frac{R^2}{0.48\,\, 10^{-8}}
\]
with $R^2 = 6.25 \,\, 10^{-12} cm^2$:
\[
\tau =7.6\,\, 10^{-3} sec.
\]
This is  the  speed  of the  transition process  with the
split-gate closed. \vskip0.5cm
\subsection{Split-gates open}
Consider the case when the barriers at  the  entrances to  the
wires are absent, ${\bf l}=0$, or the  split-gate open. In this
case the equation for resonances may be presented  as
\[
\left\{R^{-1}\left[\frac{\hat{\phi}_0\rangle\,
 \langle \hat{\phi}_0}{\hat{\lambda} - \hat{\lambda^r_0}} +
  O\left( \frac{1}{\hat{\rho}(\hat{\lambda^r_0})}\right)\right] +
   i \sqrt{\frac{m^{\parallel}}{m_0}} p\right\} e = 0,
\]
or, in non-dimensional form :
 \[
0 = \frac{|\hat{\phi}|^2}{\hat{\lambda} - \hat{\lambda^r_0}} +
O\left(4.3\right) + i\sqrt{0.9}\,\,\times \,\,250\,\,\times\,\,
0.372.
 \]
Again, we  see  that {\it the  contribution to DN-map from the
neighboring non-resonance eigenvalues is  dominated by  the main
term} : $\bigg|O\left(4.3\right)\bigg|<< \sqrt{0.9}\,\,\,250\,\,\,
0.372 = 88$. Then the non-dimensional resonance may be calculated
from the  one-pole  approximation of DN-map as
\[
\hat{\lambda} = \hat{\lambda}^r_0 - \frac{|\hat{\phi}|^2}{4.3 +
88.2 \,\, i}= \hat{\lambda}^r_0 - 5.5\,\, 10^{-3} + i\,\, 0.11,
\]
and the  life-time  of  the  resonance is  found  as
\[
\tau \approx 10^{-11} sec\,\, = \,\,10\,\,\,\, \pi s.
\]
The life-time  depends quadratically  of  the  radius of  the
well, other words,it is  quadrupled if  the radius of  the quantum
well is doubled. Vice  versa, for  the  quantum well radius $100
\,\,\, A$ the life-time  with  equivalent  other parameters  is
$\tau\approx 2 \pi s$. \vskip0.5cm

\subsection{One-pole approximation
of the  Scattering matrix}

In both  cases 4.2 and 4.3 we neglected the  contribution to
DN-map from the  non-resonance eigenvalues, using actually the
{\it one-pole approximation} for expressions staying in the
numerator and denominator of the Scattering matrix taking into
 account the  leading terms of the  DN-map only.
  Consider now the expression  for  the  Scattering matrix
combined of the leading terms  only. Then,  in the
 second  case, we obtain the  following  ``one-pole approximation" $S_{_{0}} (\lambda)$
 of  the  Scattering matrix  on the  essential interval of energy, $\lambda = p^{^2}$ :
\begin{equation}
\label{onepole} S(\lambda)\approx - \frac{\frac{P_+
\frac{\partial\varphi^r_0}{\partial n}\rangle \langle P_+
\frac{\partial\varphi^r_0}{\partial n}} {\lambda - \lambda_0} + i
\sqrt{\frac{m^{\parallel}}{m_0}}p I} {\frac{P_+
\frac{\partial\varphi^r_0}{\partial n}\rangle \langle P_+
\frac{\partial\varphi^r_0}{\partial n}} {\lambda - \lambda_0} - i
\sqrt{\frac{m^{\parallel}}{m_0}}p I} = {\bf S} (\lambda).
\end{equation}
Note  that  the  zeroes  of  the function ${\bf S} (\lambda)$ can
be  found from an elementary algebraic equation. The deviation of
them from zeroes  of the  Scattering matrix  can be estimated
rigorously based  on the estimate for  the
 non-resonance terms and the operator version of  Rouchet
theorem, see \cite{Gohberg}. One can  use the one-pole
approximation (\ref{onepole}) of the  Scattering matrix for
approximate description of the electron's transport across the
quantum well, if the  conditions (\ref{protrus1},\ref{protrus2})
are  fulfilled. \vskip1cm
\section{Switching Phenomenon}
\subsection{Shape of the resonance
eigenfunction} \noindent One  can  see  from the  last formula
(\ref{onepole}) in previous  section that the transmission  from
the input-wire $\Omega_1$  to  the  terminal $\Omega_s$ is
blocked, if for given magnitude ${\cal E}$ of the constant
electric field in the  basic domain $\Omega_0$ the projection of
the  normal derivative of the resonance eigenfunction
$\varphi^{^r}_{_0}$ of the  intermediate operator onto the
corresponding  eigenvector on the  cross-section of the  open
channel $e_{_1}^{^s} =\sqrt{\frac{2}{\delta_0}}\sin\frac{\pi
y}{\delta}$ vanishes. Practically  this  condition is  fulfilled
if a zero of the normal derivative of the resonance eigenfunction
$\varphi^{^r}_{_0}$ in the basic domain sits at the center $a_s$
of the bottom section of the wire  $\Omega_s$.  This  means  that
the single-mode transmission
 of an electron  across the quantum well is  implemented  via
excitation of  the  resonance  mode   $\varphi^{^r}_{_{0}}$ inside
the  quantum well $\Omega_0$ and  the  transmission  coefficients
are  defined  by  the  local properties  of  the  resonance
eigenfunction of the  intermediate operator in the  domain  near
to  the bottom sections $\Gamma_s$ of the  wires.
\par
It was noticed in \cite{MP01,MP02} that the design of  the network
and  the magnitude of  the constant field inside the basic domain
may be selected such that the zeroes of the normal derivative of
the resonance eigenfunction are sitting on the entrances of two
wires simultaneously, leaving the input wire and one of terminals
non-blocked.  One  can show, that  the resonance entrance vector
 $\phi^r_0 = P_+ \frac{\partial\varphi^r_0}{\partial n}$
produced from the  resonance eigenfunction  of  the intermediate
operator (with the  first  channel ``chopped off'') coincides with
the  corresponding portion  $\phi_0 = P_+
\frac{\partial\varphi_{_0}}{\partial n}$ of  the  eigenfunction
$\varphi_{_0}$ of  the  Dirichlet problem in the quantum well, see
Appendix,subsection  9.3. In this section we do not distinguish
resonance entrance vectors obtained  from eigenfunctions of the
Intermediate operator and  one of the  Dirichlet problem on the
well.
 Our  calculations  with  Dirichlet
problem in \cite{MP02} show that: if the non-dimensional amplitude
${\bf e} = \frac{2m R^3 e}{\hbar^2} \mathcal{E}$ of the
macroscopic electric field $\mathcal{E}e \langle \nu,x\rangle$
inside the quantum well $\Omega_0$ is chosen as $18.86$, then the
eigenfunction $\hat{\varphi}_0$ corresponding to the
non-dimensional \textit{resonance} eigenvalue $\lambda_{_0}=14.62$
inside the well has a single line of zeroes inside the  well and
zeroes of it's normal derivative sit on the unit circle at the
points forming angles $\pm \frac{\pi}{3} $ with the direction of
the unit vector $\nu$. This eigenfunction is even with respect to
reflection in the line spanned by the vector $\nu$. The nearest
eigenvalues of the even series of eigenfunctions sit at
$\hat{\lambda}= 2.10,\,\,25.82 $. The nearest eigenvalues of the
odd series of eigenfunctions are $5.78,\,\, 12.32,\,\, 25.99$.
\vskip0.5cm
\subsection{Transmission coefficients} The resonance
entrance vector $\hat{\phi}_0$ computed with use of the normalized
eigenfunction $\hat{\varphi}_0$ of the non-dimensional
Intermediate operator or Schr\"{o}dinger operator on the well with
the potential defined by the vector $ \nu$ directed to the point
$a_1$ (to  the entrance of the  input wire ) has the components
\cite{Robert}:
\begin{equation}  \label{entrance}
\hat{\phi}_{0} = (1,\,0.1,\,3,\,0.1),
\end{equation}
hence $||\hat{\phi}_{_0} ||^{^2} = C \approx 10$. Then the
transmission coefficients may be calculated  from the   one-pole
approximation (\ref{onepole}) as:
\[
|T_{12}|= |T_{14}|= 0.02,\,\, |T_{13}|= 0.6.
\]
Really, using the  one-pole  approximation for  the  Scattering
matrix  presented as a  function of the  geometric  spectral
parameter $\lambda = p^{^2}$ near to resonance $\lambda_{_0}$:
\[
S(\lambda) \approx - \frac{\frac{\hat{\phi}_{_{0}}\rangle\,\langle
\hat{\phi}_{_{0}} }{\lambda-\lambda_{_0}} + i p
\sqrt{\frac{m^{^{\parallel}}}{m_{_0}}}}{\frac{\hat{\phi}_{_{0}}\rangle\,\langle
\hat{\phi}_{_{0}} }{\lambda-\lambda_{_0}} + i p
\sqrt{\frac{m^{^{\parallel}}}{m_{_0}}}} = I - 2
\frac{\hat{\phi}_{_{0}}\rangle\,\langle \hat{\phi}_{_{0}}
}{|\hat{\phi}_{_{0}}|^{^2}} \frac{1}{1 +
i\sqrt{\frac{m^{^{\parallel}}}{m_{_0}}}\,\frac{p (\lambda_{_0} -
\lambda ) }{|\hat{\phi}_{_{0}}|^{^2}}},
\]
which gives the  transmission coefficients as non-diagonal
elements of the  Scattering matrix and implies the announced
result at $\lambda = \lambda_{_0}$.
 This permits to calculate the ratio of amplitudes of the
signal in closed and open wires as $1:30$ and calculate the
conductance from the input wire to the open wire $\Omega_3$ just
from the Landauer formula, see \cite{Landauer70}, since other
wires $\Omega_{_{2,4}}$ are closed.:
\begin{equation}
\label{Landauer} \sigma_{13}\approx
\frac{e^2}{h}\frac{T^{^2}_{_{13}}}{1 - T^{^2}_{_{13}}} =
\frac{e^2}{h} \frac{0.36}{0.64}.
\end{equation}
This  result holds  for  zero absolute temperature and for
spin-polarized  electrons. For  non-polarized electrons the result
should be doubled. The transmission coefficient at the resonance
energy for non-zero absolute temperature should be obtained via
averaging over the Fermi-distribution on the essential interval of
energy $(E_F - \kappa T,\, E_F + \kappa T)$, similarly to
\cite{Aver_Xu01}, and may give a result close to the previous one
(\ref{Landauer}), or close to zero in two limit  cases
\begin{equation}  \label{temper}
\kappa T << \frac{\hbar}{\tau}\,\,\mbox{or}\,\,\kappa T >>
\frac{\hbar}{\tau}
\end{equation}
respectively.
\par
The above formulae show that in certain range of temperatures the
transmission is proportional to the product of components $\langle
\frac{\partial \varphi_0}{\partial n_s}, e_s \rangle$ of the
resonance entrance vector on the bottom sections of the
corresponding wires, in complete  agreement with the basic
observation in \cite{Opening} quoted in  the  Introduction.
Similar fact  for the switch based on the quantum well with
Neumann boundary conditions was  noticed in \cite{MP01}. An analog
of it remains true for the scattering on the quantum ring, see
\cite{MP00}.

One may obviously construct the \textit{dyadic}(one input and two
terminals) Resonance Quantum Switch (RQS-2) based on the above
observation concerning the transmission coefficients.
\textit{Triadic} (three-terminal) switch (RQS-3) can be
constructed when selecting the magnitude of the governing electric
field as proposed in \cite{MP01,MP02} with the resonance
eigenfunction possessing two zeroes of the normal derivative
dividing the boundary of the well in ratio $1:2$. Taking into
account that the zeroes move on the boundary of the well together
with rotation of the vector $\nu$, one can see that the directing
the vector opposite to the contact point $a_s$ shifts the zeroes
of the normal derivative to the complementary contact points, thus
blocking the complementary wires on the essential
 interval of energy. \vskip0.5cm

\section{Working point of the triadic
Resonance Quantum Switch}

The working point $R,\mathcal{E}, V_0$ of the  triadic
 Resonance Quantum Switch (RQS) is
defined by  the position of  the resonance energy level of the
intermediate  operator (or the Schr\"{o}dinger operator inside the
well) closest to the Fermi level on the  wires, the  above
``geometric'' property of the resonance eigenfunction $\varphi_0$,
the single-mode condition on the wires, and  the temperature.  It
was
 noticed above that  \textit{the position of the working point  of the
triadic switch RQS-3 can't be defined experimentally} just by the
straightforward scanning on one of parameters with other
parameters fixed at random, since the probability of proper choice
of  remaining parameters is zero, being proportional to the
zero-measure of a point on a $2-d$ plane.

Consider the  three-terminal Resonance Quantum Switch
 (RQS) constructed in form of a circular quantum well $\Omega_{_0}$
with four quantum wires  $\Omega_{_s},\, s= 1,2,3,4,$ attached to
it at the contact points $a_1,\, a_2,\, a_3,\, a_4 ,$ selected as
suggested above, such that the  entrances to  terminals
$a_2,a_3,a_4$ divide the boundary of the circular well into three
equal parts. Assuming that the spectral  properties of the
non-dimensional Schr\"{o}dinger operator  are  already defined to
fulfill the above geometric condition on the  resonance
eigenfunction, we choose the working point of the switch in
dependence of desired temperature. Consider first the
\textit{non-dimensional} Schr\"{o}dinger equation in the unit disc
$| \xi| < 1$ with Dirichlet boundary conditions at the boundary.
It was obtained , see (\ref{ndimSch}),
 from the original equation by scaling $
x = R \xi$ in the  geometrical form  (\ref{SchrOp}) of the
Schr\"{o}dinger equation for  the  values of  energy on the
 first spectral band :
\begin{equation}
\label{connect} -\bigtriangleup_{\xi} u + \frac{2 m_0 e
\mathcal{E}R^3}{\hbar^2}\langle \xi ,\, \nu\rangle u  +
 \frac{2m_0 R^2}{\hbar^2} \left[V_0 - V_{\infty}-
\frac{\hbar^2}{2 m^{\bot} }\left(\frac{\pi}{\delta}\right)^2
\right]u = \hat{p}^2 =\hat{\lambda} u.
\end{equation}
Here $\hat{\lambda} =\hat{p}^2 =  R^2 p^2 = R^2 \lambda$ is the
non-dimensional spectral parameter,\, $\mathcal{E}$ is the
magnitude of the selected electric field, $V_0$ is  the additional
constant ``background'' potential on the well  $\Omega_0$, and
$V_{\infty}$ is  the potential in quantum wires . The unit vector
$ \nu$ defines the direction of  the  electric field, $e$ is the
 electron charge  and $R$ is the radius of the circular well. Selecting
  ${\bf e} = -\frac{2m_0\,\,e \mathcal{E}R^3}{\hbar^2} = 18.86$,
 where $e$  is {\it now} the  absolute  value of the  electron's  charge,
one may see that the eigenfunction of the  equation
\begin{equation}
\label{dimless} -\bigtriangleup_{_{\xi}} u - {\bf e} \langle
\xi,\nu\rangle u = \tilde{\lambda} u = \tilde{p}^{^{2}} u,
\end{equation}
corresponding to the second lowest eigenvalue  $14.62$
\[
\tilde{\lambda}_{_{0}} = \hat{\lambda}_{_0} - \frac{2m_0
R^2}{\hbar^2} \left[V_0 - V_{\infty}- \frac{\hbar^2}{2 m^{\bot}
}\left(\frac{\pi}{\delta}\right)^2 \right] := \hat{\lambda}_{_0} -
\hat{\bf V}_{_0}
\]
 of the even series has  proper
positions of zeroes of the normal derivative on the boundary of
the well.

The minimal distance $\hat{\rho}_0$ of $\tilde{\lambda}_0 = 14.62$
 from the nearest non-dimensional eigenvalue  $12.32$ (non-dimensional spacing) is $2.3$. The
working regime of the switch will be stable if the bound states in
the well corresponding to the neighboring eigenvalues will not be
excited at the temperature $T$ :
\begin{equation}  \label{Temper}
\displaystyle \kappa T \,\,\, \frac{2m_0 R^2}{\hbar^2} \leq
\frac{\hat{\rho}_0}{2}.
\end{equation}
This condition may be formulated in terms of the \textit{scaled
temperature } $\displaystyle \hat{T} = \frac{2m_0\,R^2
T}{\hbar^2}$ as
\begin{equation}  \label{reduced}
\kappa\hat{T}< \frac{\hat{\rho}_0}{2}= \frac{2.3}{2}
\end{equation}
The temperature which fulfills the above condition we may  call
\textit{low} temperature, for given device. If the radius $R$ of
the corresponding quantum well is small enough, then   the
condition (\ref {reduced}) can be fulfilled for some (absolutely)
high temperature, which corresponds to the relative \textit{low}
scaled temperature. For instance, the effective  mass $m_0$ of
electron {\it in the  well} ( for  a narrow-gap semiconductor )
may be small, see \cite{Adil1,Adil2}, then even the
room-temperature may be  ``low'' enough after proper scaling.
\par
  Importance of
developing technologies of producing devices of small size with
rather high potential barriers is systematically underlined when
discussing the prospects of nano-electronics, see for instance
\cite{Compano}. Use of  narrow-gap materials may  open a  way to
room-temperature devices  of  reasonably  large  dimensions.
\par
 Assume  that  the  Fermi level sits  in the  middle of the
 first  spectral band. Then we obtain the estimate
of the radius $R$ of the domain  and  the  width  of  the  wires
from (\ref {Temper}) as:
\begin{equation}  \label{R}
R^2 \leq \frac{2.3}{2 \kappa T} \frac{\hbar^2}{2 m_0},\,\,\,\,\,
\delta < \frac{R}{2}
\end{equation}
For the fixed radius $R$, the shift potential $V_0$ on the  well
$\Omega_0$ may be defined from the condition
\[
\frac{2m_0 R^2 [E_{_F} - V_0 - V_{\infty} -
\frac{\hbar^2}{2m^{\bot}}\,\,\frac{\pi^2}{\delta^2}]}{\hbar^2} =
\hat{\lambda}_0.
\]
For instance, if we choose the radius $R$ of the domain as
$\displaystyle R^2 = \frac{\hat{\rho}_{_0} \,\,\hbar^2}{4 m_{_0}
\kappa T}$ and $\delta = R/2$, we obtain the  value of the  shift
potential on the  well:
\[
V_0 = E_f  - V_{\infty} - \frac{\hbar^2 \pi^2}{2 m^{\bot}
\delta^2} - \frac{\hbar^2 \hat{\lambda}_0 }{2 m_0 R^2} =  \kappa T
\left( 60 \frac{m_{_0}}{m^{^{\bot}}} - 13\right)
\]
Finally, the electric field $\mathcal{E}$ may be found from the
condition
\[
{\bf e} = 18.86 = e \mathcal{E}\frac{2m_0 R^3}{\hbar^2},
\]
where $e$ is the absolute value of the electron charge. Hence for
the value of $R$ selected  as above we have :
\[
e\mathcal{E} R = 18.86 \frac{\hbar^2 }{2 m_0 R^2}\approx 17 \kappa
T.
\]
 The  electric  field  obtained  from this
condition is strong  enough  to  guarantee  proper  shape of the
resonance  wave-function, but not yet destructive  for
semiconductors listed below.
\par
One  can see from the  above calculations that the switch will
 work even at room temperature if the radius
$R$ of the quantum well is small enough and the geometric details
are exact.

Calculation of the radius of the quantum well (in Angstroms) for
different materials gives the following results, \cite{Shur}:

\bigskip

\begin{tabular}{llll}
Material & m/m$_{0}$ & R$_{300K}$ & R$_{77K}$ \\
Cd$_{0.15}$Hg$_{0.85}$Te & 0.0069 & 160 & 310  \\
InSb & 0.013 & 110 & 230 \\
InAs & 0.023 & 90 & 170  \\
GaAs & 0.067 & 50 & 100 \\
Si & 0.8 & 10 & 25
\end{tabular}
\vskip0.5cm \noindent The De-Broglie wavelengths of that materials
are, correspondingly, for  Nitrogen temperature, $1300 A,\,970
A,\,730 A,\,430 ,\, 110 A$ . To  obtain the above  data we  use in
the  formula  (\ref{R}) instead of  $m_0$ the average  value of
the  effective  mass
 $m$ inside the  quantum well. For Si  the  local values may
essentially deviate from the  average  value  $0.8$, depending on
local positions  of  valleys. Note  that the Fermi  surface of the
narrow-gap  material  Cd$_{0.15}$Hg$_{0.85}$Te in  momentum space
is spherically  symmetric, hence $m^{\bot}= m^{\parallel}$. This
important fact permits to  simplify  the  above  formulae for the
transmission coefficients. \vskip0.5cm
\section{Solvable model}
The above  one-pole  approximation (\ref{onepole}) and  even
 similar  ``few-poles'' approximation  (\ref{Sapprox})
 of the  Scattering matrix  of the  switch:
\[
\frac{i p I + \Lambda^{^r}_{_T} }{i p I - \Lambda^{^r}_{_T} } :=
{\bf S}_{_{T}} (\lambda),
\]
with rational essential DN-map  of the  intermediate operator
presented as a  function of the geometric spectral parameter
$\lambda = p^{^2}$:
 \begin{equation}
 \label{essential}
\Lambda^{^r}_{_T} (\lambda) = \sum_{_{\lambda_{_l}\in
\Delta_{_T}}}
 \frac{P_{_+}\frac{\partial
\varphi_{_l}}{\partial n}\rangle \langle P_{_+}\frac{\partial
\varphi_{_l}}{\partial n}}{\lambda - \lambda_l}
\end{equation}
can be interpreted as  a  Scattering matrix  of  some  solvable
model. We sketch here  the  construction of the solvable  model of
the  switch, or, more  generally,of a  multiple splitting of the
quantum wire, based on operator extension procedure, see the
general outline of extension theory  with  applications  to
general quantum solvable models in
\cite{Albeverio,Extensions,Kurasov}, and  especially  to
modelling of  Quantum networks in \cite{QSW}.
\par
We  will  construct the  model based on  ``zero-range potential
with inner
 structure'' as  an  extension of an orthogonal  sum  of
  a one-dimensional  matrix Schr\"{o}dinger with zero potential
 on the  space   $L_{_{2}} (R_{_+},\,
E_{_+})$  of  open channels
\[
l_{+} u = - \frac{d^{^2} u}{d x^{^2}} =
\frac{m^{^{\parallel}}}{m_{_{0}}}\lambda u,
\]
with respect to  the  non-dimensional coordinates  $\hat{x} = x$
and  non-dimensional spectral parameter $\hat{\lambda} = \lambda =
p^{^2}$, and some  finite-dimensional  hermitian  Hamiltonian
$H_{_{0}}$ of inner degrees of freedom (``inner  Hamiltonian"),
which  defines  the  ``inner
 structure'' of the  model. The  integration by parts on functions with
no boundary conditions at the contact point $x = 0$ gives the
boundary form of the exterior part
\begin{equation}
\label{bformext} {\cal J}_{_0} (\psi,\, \varphi) =  \langle
l_{_{+}} \psi,\,\varphi\rangle - \langle \psi ,\, l_{_{+}}
\varphi\rangle = \langle \psi',\, \varphi\rangle  - \langle
\varphi',\, \psi\rangle
\end{equation}
where  $\psi = \psi (0) = \left(\psi_{_1} (0),\,\psi_{_2} (0),
 \psi_{_3} (0), \psi_{_4}(0)
\right)$ and  the  derivatives  $\psi_{_{s}}'$ at  the node $x =
0$ are  taken in the outgoing direction. The corresponding
boundary  form for  properly  modified   inner  hamiltonian $
H_{_{0}}$ can be obtained via  abstract integration by parts, see
\cite{Extensions}:
\[
{\cal J}_{_1} (\psi^{^{in}},\, \varphi^{^{in}}) = \langle
\xi_{_{+}}^{^{\psi}},\,\xi_{_{-}}^{^{\varphi}}\rangle - \langle
\xi_{_{-}}^{^{\psi}},\,\xi_{_{+}}^{^{\varphi}}\rangle.
\]
The sum of boundary forms ${\cal J}_{_0} (\psi,\, \varphi) + {\cal
J}_{_1} (\psi^{^{in}},\, \varphi^{^{in}}) $ vanishes on the
Lagrangian plane which can  be  defined by the boundary condition
with some  hermitian matrix  $B$:
\begin{equation}
\label{bcondsw} \left(
\begin{array}{c}
\psi'\\
-\xi_{_{+}}^{^{\psi}}
\end{array}
\right) = \left(
\begin{array}{cc}
\beta_{_{00}} & \beta_{_{01}}\\
\beta_{_{10}}& 0
\end{array}
\right) \,\, \left(
\begin{array}{c}
\psi\\
-\xi_{_{-}}^{^{\psi}}
\end{array}
\right),
\end{equation}
 where  $\beta_{_{01}} = \left(\beta_{_{01}}^{^{1}},\,
\beta_{_{01}}^{^{2}}\dots \right)$  is  the  operator  connecting
the  inner  deficiency  subspace   $N_{_{i}}$ of the  modified
inner  Hamiltonian  with the entrance subspace $E_{_{+}}$ of the
open channels. This  boundary  condition  defines a  self-adjoint
operator ${\cal L}^{^{\beta}}$ which may serve a solvable model of
the Schr\"{o}dinger operator ${\cal L}$ on the quantum  switch, if
we select  the  parameters  $H_{_{0}}$ and $B$  to  fulfill
certain conditions for given  Fermi level $E_{_{F}} =
\frac{\hbar^{^{2}}} {2m_{_0}} \lambda_{_{F}} + V_{_{\infty}} +
\frac{\hbar^{^{2}}}{2m_{_{\bot}}}\,\,\frac{\pi^{^{2}}}{\delta^{^{2}}}$
from the first spectral band , $0 < \lambda_{_{F}}<
\frac{m_{_0}}{m^{^{\bot}}}\,\, \frac{3 \pi^{^2}\
R^{^2}}{\delta^{^2}}$ in the  wires and given  temperature.
\[
\lambda_{_{F}} - \frac{2 m_{_{0}}\, R^{^2}}{\hbar^{^2}} \,\,\kappa
T < \lambda_{_{s}} < \lambda_{_{F}} + \frac{2 m_{_{0}}\,
R^{^2}}{\hbar^{^2}} \,\,\kappa T.
\]
We  assume that  the  few-pole  approximation  ${S}_{_{T}}$ is
defined by  the  corresponding essential  DN-map  which is  a  sum
of  few terms  with  poles  at the  positive eigenvalues
$\lambda_{_{s}},\,\, s = 0,1,2,..n_{_{T}}$ in the corresponding
essential interval $\Delta_{_{T}} = \left(\lambda_{_{F}} - \frac{2
m_{_{0}}\, R^{^2}}{\hbar^{^2}} \,\,\kappa T,\, \lambda_{_{F}} +
\frac{2 m_{_{0}}\, R^{^2}}{\hbar^{^2}} \,\,\kappa T\right) $ of
the  spectral parameter.
\par
Scattering matrix of the  model can be  calculated in explicit
form based  on the  boundary condition (\ref{bcondsw}) connecting
the boundary data $\xi_{_{\pm}}$ of the inner component of the
wave function of the switch. Using the  equation  $ \xi_{_{-}} = -
{\cal M}\, \xi_{_{+}}$ derived  with use of the Krein's Q-function
${\cal M} (\lambda) = P_{_{N_{i}}}\frac{I + \lambda H}{H -\lambda
I}P_{_{N_{i}}}$, see \cite{Extensions},  we can eliminate the
boundary data $\xi_{_{\pm}}^{^{\psi}}$  of the inner component of
the wave  function
\[
\xi_{_{+}}^{^{\psi}} = -\beta_{_{10}} \left[ (I + S) e + s
e\right]
\]
and  obtain an expression  for ${\bf S}$ from the  matching
conditions (\ref{bcondsw}) of the inner component of the
wave-function of the  model with the  Scattering Ansatz  in the
open channels
\[
\psi (x) = e^{^{-i p x}}e  + e^{^{+i p x}} {\bf S}(\lambda) e,
\]
where  $e \in E_{_{+}}$ and ${\bf S}(\lambda)$. The  Scattering
matrix of the  model is  found  as :
\[
{\bf S}(\lambda) = \frac{ip +\left[\beta_{_{00}} - \beta_{_{01}}
{\cal M} \beta_{_{10}} \right]}{ip -\left[\beta_{_{00}} -
\beta_{_{01}} {\cal M} \beta_{_{10}} \right]}.
\]
Our  aim is: to define the  parameters $B,H$ of the  model such
that the above Scattering matrix ${\bf S}$ coincides with  the
essential
 Scattering matrix  ${\bf S}_{_{T}} (\lambda)$ of the  switch.
 \par
Denote by $Q_{_{t}}$ the spectral projection onto the eigen-space
of $H$ corresponding to the eigenvalue $\lambda_{_t}$ framed by
projections onto the deficiency subspace $N_i$
\[
Q_{_{t}} = P_{_{i}} e_{_{st}}\rangle \langle  P_{_{i}} e_{_{t}}.
\]
Then the expression  in the  numerator of the  model Scattering
 matrix takes the form :
\begin{equation}
\label{KreinQ}
 \left[ \beta_{_{00}} - \beta_{_{01}} {\cal M}
\beta_{_{10}} \right]=
 \left[\beta_{_{00}} + \sum_{_{t}}
\lambda_{_t}\beta_{_{01}} Q_{_{t}}\beta_{_{10}}\right] + \sum_{_t}
\frac{1 + \lambda^{^2}_{_{s}}}{{\lambda_{_{t}}} - \lambda}
\beta_{_{01}} Q_{_{t}}\beta_{_{10}}.
\end{equation}
We will define the boundary parameters later, but once they are
defined, we choose $\beta_{_{00}}$ such that the first summand in
(\ref{KreinQ}) vanishes: $\beta_{_{00}} + \sum_{_{t}}
\lambda_{_t}\,\beta_{_{01}} Q_{_{t}}\beta_{_{10}} = 0$. This
condition  guarantees correct  behavior of the model Scattering
matrix  for  large $\lambda$ which is  typical for  real
Scattering matrices and  for above  few-pole  approximations
(\ref{essential}) of the  DN-map $\Lambda_{_T}^{^r}$.
 Compare now the  remaining summand  $\sum_{_t}
\frac{1 + \lambda^{^2}_{_{s}}}{{\lambda_{_{t}}} - \lambda}
\beta_{_{01}} Q_{_{t}}\beta_{_{10}}$ of the  Krein- function  with
the  essential  part ${\Lambda}_{_{T}}^{^r}$ of the $DN$-map  of
the intermediate operator which is presented as a  sum of polar
terms
\begin{equation}
\label{polarterm} - \sum_{_{\lambda_{_t} \in
\Delta_{_T}}}\frac{P_{_{+}}\frac{\partial \varphi_{_{t}}}{\partial
n}\rangle \langle P_{_{+}}\frac{\partial \varphi_{_{t}}}{\partial
n}}{\lambda_{_t} - \lambda}.
\end{equation}
From (\ref{KreinQ},\ref{polarterm} ) we see, that the  eigenvalues
$\lambda_{_{s}}$ of  the model inner  Hamiltonian should  coincide
with essential eigenvalues  $\lambda_s$ of the intermediate
operator. The  boundary  parameters  and  the deficiency subspaces
of the  model should be   selected  such that all residues  of the
Krein function  coincide  with  residues of the essential part
$\Lambda_{_{T}}^{^r}$ of the DN-map of the  intermediate operator.
Then the  model Scattering matrix takes the form:
\begin{equation}
\label{ModelSmatr} {\bf S}(k) = \frac{ip I- \sum_{_s} \frac{1 +
\lambda_{_{s}}^{^2}}{\lambda_{_s} - \lambda} \beta_{_{01}}
Q_{_{s}}\beta_{_{10}}}{ip I+ \sum_{_s} \frac{1 +
\lambda_{_{s}}^{^2}}{\lambda_{_s} - \lambda} \beta_{_{01}}
Q_{_{s}}\beta_{_{10}}},
\end{equation}
with $I = I_{_+}$ equal to the  unit  operator in $E_{_{+}}$. In
case when only one resonance eigen-value $\lambda_{_0}$ of the
intermediate operator sits  on the admissible interval of energy,
the
 model Scattering matrix
\begin{equation}
\label{one_pole} {\bf S}(k) = \frac{ipI- \frac{1 +
\lambda_{_0}^{^2}}{\lambda_{_0} - \lambda} \beta_{_{01}}
Q_{_{0}}\beta_{_{10}}}{ipI +  \frac{1 +
\lambda_{_0}^{^2}}{\lambda_{_0} - \lambda} \beta_{_{01}}
Q_{_{0}}\beta_{_{10}}}
\end{equation}
is  a one-pole approximation of  the Scattering matrix  of the
network.
  \vskip1cm

\section{Conclusion}
Working parameters  of the  switch were estimated  based on
one-pole  approximation  of the  Scattering matrix which coincides
with an exact Scattering matrix of some  solvable model  of the
Quantum Switch. The developed approach, based on the observation
from \cite{Opening} quoted above in Introduction, can be used not
only for design of devices aimed to manipulation the current
across quantum dots with constant electric field, but also to
corresponding non-stationary problems, like Quantum pumping, see
\cite{Cohen03,MuzAd} and even for spin-filtering \cite{Sarma},
based  on  Rashba spin-orbital Hamiltonian,
\cite{Rashba,ByRashba84}. On the  other hand, taking into account
details of the  shape  of the wave-function encoded in the
corresponding Dirichlet-to-Neumann  map, rather than just in the
``overlapping integrals", may help understanding functioning of
molecular mechanisms. In particular, it can be  used  for
optimization of construction of  molecular mechanisms and
estimation of  admissible  errors  in design and  choice of
physical parameters of them.
\par
Note that the general program of replacement of the partial
Schr\"{o}dinger equation on complex quasi-one-dimensional
structures with properly chosen solvable models of Schr\"{o}dinger
operators is  being developed  already in a series of papers and
books
\cite{GP,Albeverio,Extensions,Novikov,Schat96,Schrader,Kurasov,
KZ01,RS01,KZ02} and  in  our papers \cite{MP00,MP01,MPP02} where
solvable models of switches were considered. \vskip0.5cm
\section{Acknowledgment}
\par
Professor G. Metakides and Doctor R.Compano (Industrial Department
of the European Commission) formulated in 1998 the question on
mathematical possibility of designing of a triadic Quantum Switch.
This problem appeared as a Work-Package in the EC-project, joint
with Solvay Institutes, Brussels: "New technologies for narrow-gap
semiconductors" ( ESPRIT-28890 NTCONGS, 1998 - 1999). A.M., B.P.,
A.Y are grateful to European Commission for formulating the
challenging question and financial support, and  to Solvay
Institute for  fruitful collaboration.
\par
The  authors are  grateful to  the  referee  who attracted  their
attention to  the  papers  \cite{Alamo,KZ02,RS01,Schat96} and  to
Professor P. Kuchment who  supplied  them the  manuscript of the
paper \cite{KZ01}.
\par
A.M and B.P acknowledge  support the Russian Academy of Sciences,
Grant RFBR  03-01- 00090. B.P. recognizes a support from the staff
research grant of the University of Auckland, 2001. A.Y.
acknowledges support from the ``Universities-of-Russia''- Program
of the  Ministry of Higher Education of Russia, Grant
UR.06.01.015.  B.P. is  grateful for hospitality to the Solid
State Computer Laboratory of Aizu University, Japan, where he
worked on the  final version of the paper with support from JSPS,
and especially to the  head  of  the laboratory Professor V.
Ryzhii for  the fruitful  discussion and Doctor V. Vjurkov for
important references and  supplying of printed materials, in
particular for reprint of the recent paper \cite{CsXu02}.
\par
The authors are grateful to Doctor Colin Fox who noticed that the
lines of zeroes of the resonance wave-function of the homogeneous
Dirichlet problem meet the boundary at the points different from
zeroes of the normal derivative. The authors are grateful also to
Mr. Kieran Robert and Doctor V. Olejnik for estimation of
components of the resonance entrance vector and  shifts of  the
resonance eigenvalue caused by attaching the  closed  channels  to
the quantum well.

\section{Appendix}

In this  section we review the  techniques of  matching solutions
of  the  Schr\"{o}dinger equation in  composite  domains and, in
particular, derive the  exact (\ref{scattmatr}) and approximate
(\ref{onepole}) expressions for the Scattering matrix used above.
Mathematical nature of some statements presented in this  section
forces us to choose for them the standard mathematical formulation
of results. More mathematical details on a single-mode resonance
scattering in Quantum Network can be found  in \cite{QSW}.
\vskip0.5cm
\subsection{Geometric and non-dimensional
 forms of the  basic  equation }

In above text we used two convenient forms of the Schr\"{o}dinger
equation: the geometrical form and the non-dimensional form. We
begin with descriptions of these forms, both  in case when the
split - gates are  present on the initial part  of the  wires
$(-{\bf l},\, 0)$, or absent, $ {\bf l} = 0$.
\par
In both cases the Schr\"{o}dinger equation on the  wires admits
separation of
 variables  when presenting  the  solution as  an expansion over
eigenfunctions  of the  cross-sections $e_l (y)=
\sqrt{\frac{2}{\delta}}\sin \frac{\pi l y}{\delta},\,\,
0<y<\delta$, $u= \sum_{l =1}^{\infty}u_l(x) e_l (y)$. Denoting  by
$V_1 (x)$ the potential  on the  wire
\[
V_1 (x) = \left\{
\begin{array}{ccc}
V_{barrier}  + V_{_\infty},&\mbox{if}& (-{\bf l}< x<0)\\
V_{_\infty}, &\mbox{if}& (0< x<\infty),
 \end{array}
 \right.
\]
we  may  present the   equations  for  the  amplitudes $u_l,\, l=
1,2,\dots $ as
\begin{equation}
\label{separ}
 -\frac{d^2 u_l}{d x^2} + \frac{2
m^{\parallel}}{\hbar^2}\left[ V_1 (x) - V_{\infty}\right] u_l -
\frac{ m^{\parallel}}{m^{\bot}} \frac{\pi^2 (l^2 - 1)}{\delta^2}
u_l =
 \frac{ m^{\parallel}}{m_{0}}
 \lambda u_l , \,\, l= 1,2,\dots.
\end{equation}
with the spectral parameter $\lambda = p^2 = \frac{2 m_0}{\hbar^2}
 \left[E-V_{\infty}- \frac{\hbar^2}{2m^{\bot}}
 \frac{\pi^2}{\delta^2} \right]$. Here $p$\, is the ``effective
wave-number '' in the  wires. The function $u = \sum_{l
=1}^{\infty}u_l(x) e_l (y)$ should match, on the sum $\Gamma =
cup_{s} \Gamma_s$ of the bottom sections $\Gamma_s,\, s=
1,2,\dots$ of the wires, the corresponding solution $u_0$ of the
Schr\"{o}dinger equation with linear potential on the  well. This
equation  may be  presented  in geometrical form as
\[
 \frac{2m_0}{\hbar^2}l u_0 =
  -\bigtriangleup u_0 +  \frac{2m_0}{\hbar^2}
  \left[{\cal E}e\langle x,\nu \rangle + V_0 - V_{\infty}
  - \frac{\hbar^2}{2m^{\bot}}
 \frac{\pi^2}{\delta^2}\right]u_0 =
 \]
 \begin{equation}
\label{SchrOp}
   \frac{2m_0}{\hbar^2} \left[E - V_{\infty} - \frac{\hbar^2}{2m^{\bot}}
 \frac{\pi^2}{\delta^2} \right] u_0 = \lambda u_0,
\end{equation}
\begin{equation}
\label{match} \left[u - u_0\right]\bigg|_{\Gamma_s}= 0,\,\,
\left[\frac{1}{m^{\parallel}}\frac{\partial u}{\partial n} -
\frac{1}{m_0} \frac{\partial u_0}{\partial
n}\right]\bigg|_{\Gamma_s}= 0.
 \end{equation}
In  terms of the spectral parameter  $\lambda = p^{^2}$ we  can
re-write  the  above  equation  (\ref{SchrOp}) in geometric form
with use of  the re-normalized  shift potential  ${\bf V}_{_{0}} =
\frac{2m_0}{\hbar^2}
  \left[ V_0 - V_{_\infty}
  - \frac{\hbar^2}{2m^{^\bot}} \frac{\pi^{^2}}{\delta^{^2}}\right]$
  and  the  re-normalized  electric potential
${\bf e} \langle x,\, \nu \rangle := - \frac{2m_0}{\hbar^{^2}}
  {\cal E}e\langle x,\nu \rangle $ :
  \begin{equation}
  \label{welleq}
-\bigtriangleup u_0 - {\bf e} \langle x,\, \nu \rangle  u_0 + {\bf
V}_{_{0}} u_0 = \lambda  u_0 = p^{^2} u_{_0}.
  \end{equation}
The  Schr\"{o}dinger equation on the  wires  reduced  to the  same
spectral parameter  takes the  form
\begin{equation}
\label{wireq} - \frac{d^{^2} u_{_s}}{d x^{^2}} -
\frac{m^{^{\parallel}}}{m^{^{\bot}}} \frac{d^{^2} u_{_s}}{d
y^{^2}} - \frac{m^{^{\parallel}}}{m^{^{\bot}}}\,
\frac{\pi^{^{2}}}{\delta^{^2}}\, u_{_s} + {\bf V}_{_{s}} (x)
u_{_s} = \frac{m^{^{\parallel}}}{m_{_{0}}} \lambda u_{_{s}}
\end{equation}
 with corresponding
  effective  potential  ${\bf V}_{_{s}} (x)$ vanishing  for  $x > 0$.
\par
We  assume now that the potential in the  wires  is  constant
$V_{_{s}}(x) = V_{_{\infty}}$ ( the  barrier on the  initial part
 of each  wire is  absent, ${\bf l} = 0 $).
  As before in sections 1,2 we  present the  Schr\"{o}dinger
equation in the wires  in``geometric form"  on the open channel
as:
\begin{equation}
\label{geomwir}
  - \frac{d^2 u}{d x^2} = \left\{
\frac{2m^{\parallel}\left[E-V_{\infty}\right]}{\hbar^2} -
\frac{m^{\parallel}}{m^{\bot}}
\left(\frac{\pi^2}{\delta^2}\right)\right\} u : =
\frac{m^{\parallel}}{m_0} p^2 u = \frac{m^{\parallel}}{m_0}
\lambda u
\end{equation}
and  on the  closed  channels  $l = 2,3,\dots$ as
\[
- \frac{d^2 u}{d x^2} = -
\frac{m^{\parallel}}{m_0}\left[\left(\frac{l^2
\pi^2}{\delta^2}\right)\,\,\frac{m_0}{m^{\bot}} -
\frac{2m_0\left[E-V_{\infty}\right]}{\hbar^2}\right] u := -
\frac{m^{\parallel}}{m_0} \left[\frac{(l^2 - 1 )\pi^2}{\delta^2}
\,\,\frac{m_0}{m^{\bot}} - p^2\right]u.
\]
 The  geometric  form of  the Schr\"{o}dinger equation on the  quantum well is
the same, (\ref{welleq}). Introducing  the  non-dimensional
variable $\xi = x/R$ and the non-dimensional  coefficients
\[
{\bf e}= \frac{2 m_0 R^2{\cal E}e }{\hbar^2},\,\, \hat{\bf V}_0 =
R^2 \left[\frac{2 m_0}{\hbar^2} \left(V_0 - V_{\infty}\right)-
\frac{\pi^2}{\delta^2}\,\, \frac{m_0}{m^{\bot}}\right]
\]
we  may  present the  equations in the open channel of  the  wires
 in non-dimensional form:
\begin{equation}
\label{nondimwir}
  -\frac{d^2 \hat{u}}{d \xi^2} =\frac{m^{\parallel}}{m_0} \hat{p}^2 \hat{u}=
   \frac{m^{\parallel}}{m_0}\hat{\lambda} u
\end{equation}
with  the  non-dimensional wave-number $\hat{p} = p R$  or
non-dimensional spectral parameter $\hat{\lambda}$ in the first
channel, and in  form
\begin{equation}
\label{ndimwirup}
  -\frac{d^2 \hat{u}}{d \xi^2} +\frac{m^{^{\parallel}}}{m^{^{\bot}}}\,
   \frac{R^{^2}\pi^{^2}(l^{^2} - 1)}{\delta^{^2}} u=\frac{m^{\parallel}}{m_0} \hat{p}^2 \hat{u} =
   \frac{m^{\parallel}}{m_0}\hat{\lambda} u
\end{equation}
in  closed  channels. On the  quantum well it  takes  the  form :
\begin{equation}
\label{nondimvert}
  - \bigtriangleup_{\xi} \hat{u} - {\bf e}
\langle \xi, \nu \rangle \hat{u} + \hat{\bf V}_0 \hat{u} =
\hat{p}^2 \hat{u},
\end{equation}
with $\hat{V}_{_0} = R^{^2}\, V_{_0}$. \vskip0.5cm
\subsection{Dirichlet-to-Neumann Map}
{\bf Standard DN-map} Dirichlet-to-Neumann map for the
Schr\"{o}dinger equation in geometric form in a  domain $\Omega$
 is a map of the boundary values $u_{\Gamma}$ of the
solution
\[
-\bigtriangleup u + {\bf V} u = \lambda u,\,\, u|_{_{\partial
\Omega_0}} = u_{\Gamma}
\]
on the border  $\partial \Omega = \Gamma$ of  the  domain into the
boundary  values of it's normal derivative:
\[
\Lambda : u_{\Gamma} \to \frac{\partial u}{\partial
n}|_{_{\partial \Omega_0}}.
\]
In electrodynamics (with ${\bf V} = 0 $) this corresponds to the
connection  of the potential on the boundary with the normal
current. Detailed description of general features of the DN-map
and it's  relations  to  the Scattering Matrix may be found in
\cite{SU2,DN01}, respectively. We  will review  here only  basic
features  of the  standard DN-map.
\par
Denote by $L_{_D}$ and $L_{_N}$ the self-adjoint operators defined
in $L_2 (\Omega_0)$ by the above differential expression $L u =
-\bigtriangleup u + {\bf V} u $ and homogeneous Dirichlet and
Neumann boundary conditions respectively. Corresponding Green
functions $G_{N,D} (x,\, y,\lambda)$ and the Poisson kernel
\[
{\cal P}_{\lambda} (x,\, s) = {-}\frac{\partial G_{D} (x,\, s,
\lambda)}{\partial n_s}, \,s\in \partial \Gamma,
\]
exist if  $\lambda$ is  not an eigenvalue  of the corresponding
operator  $L_{_{D}}$ (is ``regular" ). Solutions of classical
boundary problems for operators $L_{N,D}$ may be represented for
regular $\lambda$ by the ``re-normalized" potentials  of densities
supported  by the boundary $\Gamma_0$. For instance  for the
Neumann problem
\[
L u = \lambda u,\,
\]
\[
\frac{\partial u}{\partial n}\vert_{ \Gamma} = \rho
\]
with N-regular ( regular for  Neumann problem) $\lambda$ we obtain
the solution as a  re-normalized  simple-layer potential:
\begin{equation}
\label{s_layer}
    u(x) = \int_{\Gamma}  G_{_N}
(x,\,s,\, \lambda) \rho (s) d\Gamma.
\end{equation}
For  the  Dirichlet problem with D-regular $\lambda$
\[
L u = \lambda u, \,\, u\big|_{\Gamma}= u_{_{\Gamma}}
\]
we  obtain the  solution as a re-normalized  double-layer
potential:
\begin{equation}
\label{d_layer} u (x) = \int_{ \Gamma} {\cal P}_{D}
(x,\,s,\,\lambda) u_{_{\Gamma}}(s) d\Gamma.
\end{equation}
Generally the standard DN-map  is  represented  for  regular
points $\lambda$ of $L_{D}$ as a generalized  integral operator
with a singular kernel:
\[
\left(\Lambda (\lambda) u_{_{\Gamma}}\,\right) (x_{ \Gamma} ) =
\]
\begin{equation}
\label{DNmap}
   \frac{\partial}{\partial n} \vert_{ x = x_{ \Gamma}}
     \int_{ \partial \Omega} {\cal
P}_{D} (x,\,s,\,\lambda)u_{_{\Gamma}} d\Gamma
\end{equation}
and  exist as  an  operator in proper  functional  classes. In
particular for operators defined  on  $W_2^2 (\Omega )$ it acts in
proper  Sobolev classes , see \cite{Embedding}, as :
\[
\Lambda (\lambda) : W_{_2}^{^{3/2}}(\Gamma)\longrightarrow
W_{_2}^{^{1/2}}(\Gamma)
\]
see for instance \cite{SU2,DN01}. One can see from the
straightforward integration by  parts  that the DN-map is  an
analytic  function of   the spectral parameter $\lambda$ with a
negative imaginary part (for interior problem, with an outer
positive normal on the boundary).
\par
      The  following simple statement, see \cite{DN01},
shows, that the singularities  of  the kernel of the DN-map
$\Lambda_{in} (\lambda)$ in  space variable  and  the poles  at
the eigenvalues  of  the inner Dirichlet problem  may be in
certain sense  separated :
\begin{theorem}{Let us consider the Schr\"{o}dinger operator
     $ L_{_D} = -\bigtriangleup + q(x) $ in $L_2
(\Omega)$  with  real   potential $q$  and  homogeneous  Dirichlet
boundary condition on the smooth boundary $\Gamma$ of the bounded
domain $\Omega$. Then the DN-map $\Lambda $  of  $L_{_D}$ defined
on the  Sobolev class  $W_{2,0}^2 $ satisfying homogeneous
Dirichlet
 boundary  conditions  on  $\partial{\Omega}$ has  the
following representation on the  complement  of the corresponding
spectrum $\sigma_L $ in complex  plane  $\lambda,\, M > 0$:
\begin{equation}
\begin{array}{c}
\Lambda (\lambda) = \Lambda (-M) - (\lambda + M){\cal
P}^+_{-M}{\cal P}_{-M} -
\\
(\lambda + M)^2 {\cal P}^+_{-M} R_{\lambda} {\cal P}_{-M},
\end{array}
\label{Sep}
\end{equation}
where  $R_{\lambda}$ is the  resolvent  of $L_{_{D}}$, and  ${\cal
P}_M$ is the Poisson map  of  it. The  operators
\[
\Lambda(-M),\, \left( {\cal P}^+_M {\cal P}_M\right) (x_{\Gamma},
y_{\Gamma)}
\]
are bounded in proper  Sobolev classes: respectively from
$W^{3/2}_{2} (\Gamma)$ onto $W^{1/2}_{2} (\Gamma)$ and in
$W^{3/2}_{2} (\Gamma)$, and the operator
\[
-(\lambda + M)^{^2}\,\,\left({\cal P}^+_M R_{\lambda} {\cal P}_M
\right) (x_{\Gamma}, y_{\Gamma})
\]
may be presented by the convergent spectral series
\begin{equation}
\label{DNres} -(\lambda + M)^{^2}\,\,\sum_{\lambda_s \in \sigma_L}
\frac{ \frac{\partial\varphi_s}{\partial n} (x_{\Gamma})
\frac{\partial\varphi_s}{\partial n} (y_{\Gamma})} {(\lambda_s +
M)^2  (\lambda_s -\lambda) }
\end{equation}
and is  compact  in  $L_2(\Gamma)$. }
\end{theorem}
Note that the  last term of the  sum  (\ref{Sep}) presented  as
(\ref{DNres}) reveals an  explicit dependence of the standard
DN-map of
 the  eigenvalues of the  inner  Dirichlet problem in $\Omega_{_0}$.
The  corresponding  formal  ``spectral expansion"  for  the kernel
of the  DN-map
\begin{equation}
\label{formalDN} \Lambda (\lambda) (x_{_{\Gamma}},\,
y_{_{\Gamma}}) = \sum_{\lambda_s \in \sigma_{L_D}} \frac{
\frac{\partial\varphi_s}{\partial n} (x_{\Gamma})
\frac{\partial\varphi_s}{\partial n} (y_{\Gamma})} {(\lambda_s
-\lambda) }
\end{equation}
is  divergent. \vskip0.3cm \noindent{\bf Scattering matrix via
standard matching } The standard DN-map permits to formalize the
procedure of matching solutions of partial differential equation
in the composite domain $\Omega_0 \cup \Omega_1\cup \Omega_2 \cup
\Omega_3 \cup \Omega_4$ which have a common piece of boundary
$\Gamma_1  \cup \Gamma_2 \cup \Gamma_3 \cup \Gamma_4 = \Gamma $.
\par
Denote by  $E_+$ the $4$-dimensional subspace in  $L_2 (\gamma)=
E$ spanned by the vectors $e^1_s,\, s=1,2,3,4$. $E_+$ plays  the
role of  the  ``entrance  subspace" of  open channels in the wires
if $ 0 < p^2 <
\frac{m_0}{m^{\bot}}\left(\frac{3\pi^2}{\delta^{^{2}}}\right)$.
The orthogonal complement of it $E\ominus E_+ = E_- $ is the
entrance subspace of  closed channels. On the first spectral band
\[
0 \leq \lambda \leq \frac{m_0}{m^{\bot}}\,\,
\frac{3\pi^2}{\delta^2}
\]
there are two bounded  exponential modes  of the first order based
on the  cross-section eigenfunction $e^1_s
=\sqrt{\frac{2}{\delta}}\sin \frac{\pi y}{\delta}$ in the  wire
$\Omega_s$ with  exponentials  defined  by  $p$:
\[
f_s^{\pm} (x,y)=
  e^1_s e^{\pm i \sqrt{\frac{m^{\parallel}}{m_0}}p \,\,\,x} \,\,\mbox{if} \,\,x> 0,
\]
and only one bounded exponential mode order $l$ in upper channels
$l>1$
\[
f_{s}^{l}(x,y) = e^l_{s} e^{- \sqrt{\frac{m^{\parallel}}{m_0}}\,\,
 \sqrt{ \frac{m_0}{m^{\bot}}\frac{\pi^2
\left(l^2-1\right)}{\delta^2}\, -\, p^2}\,\,\, x}\,\,\,
\mbox{if}\,\,\, x> 0,
\]
on the  upper channels (spectral  bands) $l = 2,3,\dots $. The
corresponding  Scattering  Ansatz  in the wires $\Omega_s$ is
combined as
\begin{equation}
\label{Sanzatz}
  \Psi_s (x)= \delta_{s1} f_{1s}^{-} +  f_{1s}^{+} S^1_{s1} +
\sum_{l = 2}^{\infty} S^l_{s1} f^l_{s1}
\end{equation}
with coefficients $ S^l_{s1}$ to  be defined from the matching
conditions with the  corresponding  solution of  the above
Schr\"{o}dinger   equation  inside  the quantum well $\Omega_0$.
Choosing the  outside positive normal on the boundary $\Gamma =
\partial \Omega_0$, we can present the matching
conditions with use of the  standard
 {\it Dirichlet-to-Neumann map}
(DN-map) of the  quantum well $\Omega$. The above Ansatz
(\ref{Sanzatz}) is already decomposed
\[
  \Psi_s = \left[\delta_{s1}
f^-_1 + S^1_{s1} f^+_1\right] + \sum_{l= 2}^{\infty} S^l_{s1}
f^l_s = \Psi_s^+ + \Psi_s^-,
\]
as  a  sum  of  vectors  from  $E_{\pm}$ respectively. Our aim is
: to find the  coefficients  $S^1_{st}$ of  the  Ansatz which form
the Scattering  matrix. We  may  find  them from  the condition
(\ref{match1}) of  continuation of  the  Scattering  Ansatz inside
the  domain. Denoting  by  $\Psi_{\gamma},\frac{\partial
\Psi_\gamma}{\partial n_{\gamma}}$ the  boundary  data of  the
above Scattering Ansatz on $\Gamma$, and assuming that the
boundary values of the component $\Psi_{_0}$ of the  scattered
wave  inside the  quantum well coincide on bottom sections of
wires  with the boundary  values of the Scattering Ansatz we may
present these conditions with DN-map
  of  the  Schr\"{o}dinger operator in  $\Omega_0$:
\begin{equation}
\label{matching}
 \frac{m_0}{m^{\parallel}} \,\,\frac{\partial \Psi_{\gamma}}{\partial
n_{\gamma}} =
 \frac{\partial \Psi_{_0}}{\partial n_{\gamma}}=
\Lambda^{^{^0}} \Psi_{\gamma}.
\end{equation}
Denote by  $K^+ , \bar{K}^+,\, K^- $  the operators in $E_+$ which
compute the components of the normal derivatives of the
exponential modes on the  bottom  sections  of the wires
$\Omega_s$ in the open and closed channel
\[
{K}^{+}  = i p \sqrt{\frac{m^{\parallel}}{m_0}}I,\,\, \bar{K}^{+}
= -i p \sqrt{\frac{m^{\parallel}}{m_0}}\,\, I
\]
\[
K^{-}_l = \sqrt{\frac{m^{\parallel}}{m_0}}\,\,
\sqrt{\frac{\frac{m_0}{m^{\bot}}\left(l^2 - 1 \right)
\pi^2}{\delta^2} - p^2}\,\, I,\,\,
\]
\[
 l=2,3,\dots,\,\,\, K^{-}=
\mbox{diag}\left\{K^{-}_l \right\}_{l=2}^{\infty},
\]
with positive  square  root, and  by  $P_{\pm}$ the orthogonal
projections in  $E$ onto the subspaces $E_{\pm}$.The matrices
$K^+$ are  $4\times 4$ matrices proportional to the unit matrix,
since potentials on the wires are equivalent.Then the above
equation (\ref{matching}) may be presented as a matrix equation
with respect to  the components $\Psi_s^{\pm} $ of  the  above
decomposition of $E = E_+ \oplus E_-$. Elements of the subspace
$E_+$ belong  to the Sobolev class $W^{^{3/2 - \varepsilon}}_{_2}
(\Gamma)$, hence the operators $P_+ \Lambda^{^0} P_+ ,\,P_+
\Lambda^{^0} P_-,\, P_- \Lambda^{^0} P_+ , \, P_- \Lambda^{^0}
P_-$  constructed  via  framing of the  DN-map $\Lambda^{^0}$
 of the  Schr\"{o}dinger
operator on the  Quantum by projections onto entrance  subspaces
of  open and  closed  channels
 exist as  operators in proper  classes.
 Denoting them by $\Lambda^{^0}_{++},\, \Lambda^{^0}_{+-},\,
\Lambda^{^0}_{-+},\, \Lambda^{^0}_{--}$ respectively,and  by $S^1$
the $4\times 4$ matrix with elements $S^1_{st}$ and set
$(\Psi_s^{+})_{s=1}^4(0)= \Psi^+ ,\, (\Psi_s^{-})_{s=1}^4(0)=
\sum_{l=2}^{\infty} S^l \Psi^+$.
\begin{theorem}{\it  The Scattering Matrix on the  whole  network
$\Omega_0 \cup \Omega_1 \cup \Omega_2 \cup \Omega_3 \cup \Omega_4$
 may  be  presented  in terms of  the  Dirichlet-to-Neumann map
 $\Lambda^{^{^0}}$  of  the  quantum well  $\Omega_0$ as
\begin{equation}
\label{Scatt} S(\lambda) = -  \frac{\Lambda^{^0}_{_{++}} -
\Lambda^{^0}_{_{+-}}\frac{I}{K^{^{^-}} +
\Lambda^{^0}_{_{--}}}\Lambda^{^0}_{_{-+}} -
\bar{K}^{^+}}{\Lambda^{^0}_{_{++}} -
\Lambda^{^0}_{_{+-}}\frac{I}{K^{^{^-}} +
\Lambda^{^0}_{{--}}}\Lambda^{^0}_{_{-+}} - K^{^{+}}}.
\end{equation}
The Scattering  Matrix  may be  presented as a function of the
non-dimensional spectral parameter  $\hat\lambda = R^2 \lambda $
which correspond to the scaled quantum well $\hat{\Omega}_0$
radius $1$ . The  corresponding expression may be  obtained as  a
Scattering Matrix of the  scaled network via replacement of  the
DN-map $\Lambda^{^{^0}}(\lambda)$ of the quantum well $\Omega_0$
by the  DN-map of  the  scaled  domain
 $\hat{\Lambda}^1 (\hat\lambda) = R \Lambda^{^0} (\lambda)$ and proper scaling
 of  operators :$ K^{^+} \to \hat{K}^{^+} = R K^{^+},\,\, {K}^{^{^-}} \to \hat{K}^{^{-}} = R K^{^-}$,
 reducing the  wires  width $\delta $ to  $\frac{\delta}{R}$:
\begin{equation}
\label{ScattND}
  \hat{S}(\hat{\lambda}) = -  \frac{ \hat{\Lambda}^{^0}_{_{++}} -
\hat{\Lambda}^{^0}_{_{+-}}\frac{I}{\hat{K}^{^{^-}} +
\hat{\Lambda}^{^0}_{_{--}}}\hat{\Lambda}^{^0}_{_{-+}} -
\bar{\hat{K}}^+}{\hat{\Lambda}^{^0}_{_{++}} -
\hat{\Lambda}^{^0}_{_{+-}}\frac{I}{\hat{K}^{^{^-}} +
\hat{\Lambda}^{^0}_{_{--}}}\hat{\Lambda}^{^0}_{_{-+}} -
\hat{K}^{^+}}.
\end{equation}
\par
}
\end{theorem}
{\it Proof} is  obtained  based on orthogonal decomposition of the
whole  space  $L_{_{2}} (\Gamma)$ into  orthogonal sum of  open
and  closed channels, followed by
 the  straightforward calculation,
see \cite{QSW}. \vskip0.5cm \noindent{\bf  Partial DN-map}
 The  Scattering  Matrix (\ref{ScattND}) contains
a  special  combination of  matrix elements  of  the  DN-map
$\hat{\Lambda}^{^{0}}$ of  the  quantum well
 \begin{equation}
\label{DNr} \hat{\Lambda}^{^{r}} = \hat{\Lambda}^{^{0}}_{++} -
\hat{\Lambda}^{^{0}}_{+-}\frac{I}{\hat{K}^{^{^-}} +
\hat{\Lambda}^{^{0}}_{--}}\hat{\Lambda}^{^0}_{-+}.
 \end{equation}
 This function  has  negative  imaginary part in upper half-plane
 $\Im \lambda > 0$ and  singularities  at the vector-zeroes of the  denominator
 $\left(\hat{K}^{^{^-}} + \hat{\Lambda}^{^0}_{--}\right) e = 0$.
 The following  statement, see \cite{QSW} gives an  interpretation of  this  function
 in terms of  the  intermediate  operator ${l}^r$ :
\begin{theorem}{\it  The Intermediate operator  $l^{^r}$ defined on the  whole  network
$\Omega_0 \cup \Omega_1 \cup \Omega_2 \cup \Omega_3 \cup \Omega_4
$ by  the  Schr\"{o}dinger differential expression  (\ref{SchrOp})
and the  boundary conditions (\ref{match_2}) and (\ref{cho_p}) is
self-adjoint.
 The continuous spectrum of the  non-trivial component
${l}^r$ of this operator on the  orthogonal complement of  the
first  channel consists of  the  branch $\lambda \,\geq
\,\frac{m_0}{m^{\bot}}\, \,3\,\frac{\pi^2}{\delta^2}$ with the
countable sequence of thresholds. With geometrical spectral
variable $\lambda$ the eigenvalues of  the operator ${l}^r$below
the  threshold $\frac{m_0}{m^{\bot}}\,\,3
\,\frac{\pi^2}{\delta^2}$  coincide with the zeroes  $\lambda^r_s$
of the  denominator $\left({K}^{^{^-}} +
{\Lambda}^{^0}_{--}\right)$. The DN -map of  the  operator ${l}^r$
on the  whole network with chopped-off first  channel coincides
with  the  operator-function (\ref{DNr})}.
\end{theorem}
This  statement  also can be  verified  via straightforward
calculation, see  \cite{QSW} based on  the fact that  the
intermediate operator is  self-adjoint.
 Note  that the  DN-map of the  intermediate
operator  is  actually  a  finite  matrix and  it's  calculation
requires only  finite  operations, eliminating  infinite  series:
beginning  from the  Green - function  $G^{^{r}} (x,y)$ of the
self-adjoint  intermediate operator   $l^{^{r}}$  we consider the
restriction of the  corresponding Poisson map with the kernel
 $-\frac{\partial G^{^{r}}}{\partial n_{_y}} (x,y),\,\, y\in \Gamma$
 onto  the  $4$-dimensional subspace $E_{_{+}}$ of the
 entrance vectors $\bigvee_{_{s = 1}}^{^4} e_{_s}$
 of the  open  channels. Then  anticipating the partial matching
 conditions (\ref{match1}) in the  open channel only
 we consider  the  projection of the  normal derivative
 of the  result on $\Gamma$ onto  $E_{_+}$ and  obtain the  kernel  of the  DN-map in form
 of $4\times 4$ matrix  from $E_{_{+}}\times E{_{+}}$ :
\begin{equation}
\label{DNr_ker}
\Lambda^{^r}(x_{_{\Gamma}},x_{_{\Gamma}},\,\lambda) = -
P_{_{+}}\frac{\partial^{^2} G^{^{r}}}{\partial n_{_x}\,\,\partial
n_{_y}} P_{_{+}}
\end{equation}
Note  that the  vectors from the entrance subspace $E_{_+}$ of the
open channels, being continued by zero to the complement of the
sum $\Gamma$ of bottom sections, belong to proper Sobolev classes
on the whole boundary of the network so that  the projection
$P_{_{+}}$ can be correctly calculated.
 \vskip0.5cm
\noindent {\bf Scattering matrix via partial DN-map} To calculate
the Scattering matrix in terms of the partial DN-map of
 the  Intermediate operator  we  should  match  the   restriction  of the  Scattering
 Anzatz  on the  sum $\Gamma = \cup_{_{s=1}}^{^{4}} \Gamma_{_{s}}$ of  bottom section
 of open channels  {\it only}:
 \[
\Psi_{_s}^{^{+}}= \delta_{s1} f_{1s}^{-} +  f_{1s}^{+}
S^1_{s1},\,\,s=1,2,3,4,
\]
\[
{\bf \Psi} = \left\{\Psi_{_s}^{^{+}}\right\}_{_{s=1}}^{^{4}} =
F_{_{in}}\,e_{_{1}} +
 F_{_{out}} S e_{_{1}}
 \]
with  the  Jost  matrices $F_{_{in, out}}$, to the  solution
$\Psi_{_{0}}$ of   the   Intermediate  homogeneous equation :
\begin{equation}
\label{intermatch} P_{_{+}} \Psi_{_{0}} = {\bf \Psi},\,
\frac{1}{m_{_{0}}}\, \frac{\partial \Psi_{_{0}}}{\partial
n_{_{s}}} = \frac{1}{m^{^{\parallel}}}\, \frac{\partial
\Psi_{_{s}}}{\partial n_{_{s}}},\, s= 1,2,3,4.
\end{equation}
Substituting  here  the above  expression (\ref{DNr}) for the
partial DN-map, and  using the  fact that  all  wires  are
equivalent, hence the
 Jost matrices are  proportional to the  unit matrix in open channels, we  obtain  an
equation  for  the  Scattering  matrix:
\[
\frac{1}{m_{_0}} \Lambda^{^{r}} F_{_{in}} -
\frac{1}{m^{^{\parallel}}}  F_{_{in}} = - \left( \frac{1}{m_{_0}}
\Lambda^{^{r}} F_{_{out}} - \frac{1}{m^{^{\parallel}}}
F_{_{out}}\right),
\]
which has the  solution :
\[
 S = -\frac{F_{_{in}}}{F_{_{out}}}
 \frac{\Lambda^{^r} - \frac{m_{_{0}}}{m^{^{\parallel}}} \frac{F'_{_{in}}}{F_{_{in}}}}
 {\Lambda^{^r} - \frac{m_{_{0}}}{m^{^{\parallel}}} \frac{F'_{_{out}}}{F_{_{out}}}}
\]
where  the  denominator is  preceding the  numerator, and  the
fractions $\frac{F_{_{in}}}{F_{_{out}}}=
P_{_+}\frac{F_{_{in}}}{F_{_{out}}},\, \frac{F_{_{in}}}{F_{_{out}}}
= P_{_+} \frac{F_{_{in}}}{F_{_{out}}} $ are  proportional to the
unit matrix in $E_{_{+}}$. Substituting here the  explicit
expression  for  the  Jost  solutions in  the  open channel we
obtain the  expression  (\ref{scattmatr}) for the  Scattering
matrix in terms  of  DN-map of the  intermediate operator.
\par
Note that using of  the  partial DN-map of the  intermediate
operator automatically  provides the  matching conditions in
closed  channels. \vskip0.5cm \noindent
\subsection{Compensation of singularities}
Singularities  of the Scattering matrix  sit  near the
eigenvalues of the  Intermediate operator, which are  shifted
with respect to eigenvalues of the Schr\"{o}dinger operator  with
Dirichlet boundary conditions  on the  boundary of the  well. The
shift, generically, is  not  large, but may be essential when
searching for the  working point. It may be  estimated via
standard  analytical perturbation procedure. subsection.An
important fact used above in  section  3 is :
\begin{theorem}{\it
The  pole $\lambda_0$ of  the  DN-map $\Lambda^{^{0}}$ of the
quantum well, which is the  singularity of  the  first  addendum
${\Lambda}^{^{r}}_{++}$ of (\ref{DNr}), is compensated by the pole
of the second addendum and disappears as a  singularity of  the
whole  function $\Lambda^r$ , so  that  the whole expression
(\ref{DNr}) is, generically,  regular at the point $\lambda_0$. A
new pole appears as a zero of the denominator ${K}^{^{^-}} +
{\Lambda}^{^{0}}_{--}$ and  coincides  with the eigenvalue of the
intermediate  operator. The corresponding  residue is  combined of
root vectors which correspond to this new pole and coincide, in
the first order of the perturbation procedure, with the resonance
entrance vector $P_+ \frac{\partial \varphi_0}{\partial n} =
\phi_0$ on the quantum well  $\Omega_0$.}
\end{theorem}
{\it Proof}  is  obtained in  \cite{QSW}.
 Here  we  will  verify the  corresponding non-dimensional
statement for  the  special case of the switch based on the
circular quantum well. Assume, that the Fermi level is in the
middle of the  first spectral band in the wires, $\mu_{_{1}}=
\mu_{_{2}} = 1/2$.
 Then  the effective  wave - number $p$ at the  Fermi level is
equal to $\sqrt{\frac{3 m_0}{2m^{\bot}}\frac{\pi^2}{\delta^2}}$
and  the  corresponding
 non-dimensional wave-number  $\hat{p} = Rp$ is
\begin{equation}
\label{nondimmom} \hat{p} = R p = \pi \frac{R}{\delta}
\sqrt{\frac{3 m_0}{2m^{\bot}}}\approx 8.8 \frac{R}{\delta},
\end{equation}
if  $\frac{m_0}{m^{\bot}} = 5.2 $  ( for Si). Then  the  term
 $- \bar{\hat{K}}^+$ in the
non-dimensional  expression  for  the  Scattering  matrix  may  be
estimated  as
\[
- \bar{\hat{K}}^+ = i R\,\,\sqrt{\frac{m^{\parallel}}{m_0}} p I =
i \pi \frac{R}{\delta} \sqrt{\frac{3 m^{\parallel}}{2m^{\bot}}} I
\approx i\,\, 8.8\,\, \frac{R}{\delta} I
\]
Practically  it  contains  a  ``large" parameter  compared  with
non-dimensional inverse  spacing  $|\hat\lambda_0 -
\hat\lambda_1|^{-1}= \frac{1}{2.3}$ already for $R > \delta/2$:
\[
\frac{10}{2.3}= 4.3 << 17.6.
\]
For  $R = 10 \delta$ the  corresponding inequality is $4.3 << 88$.
For the normalized  eigenvectors  $\hat{\varphi}_s$ of the
Dirichlet problem in  the modified well $\Omega_0$  radius  $1$,
with flattened  pieces of  boundary and  properly  transformed
eigenfunctions, see suggestion  above in 3.2, we  introduce the
unified  notations:
\[
\phi_{_0} = \hat{\phi}^{^{+}}_{_0} \hat{\phi}^{^{\pm}}_{_s} =
P_{\pm}\frac{\partial \hat{\varphi}_s{_s}}{\partial n},
\]
and  separate the  resonance term in  the  DN-map  framed  by
projections onto $E_{\pm}$:
\[
P_+\hat{\Lambda}P_+ = \frac{\hat{\phi}^{^{+}}_0\rangle \langle
\hat{\phi}^{^{+}}_0} {\hat{\lambda} - \hat{\lambda_0}} +
\sum_{s\neq 0} \frac{\hat{\phi}^{^{+}}_s\rangle \langle
\hat{\phi}^{^{+}}_s} {\hat{\lambda} - \hat{\lambda_s}} :=
 \frac{\hat{\phi}^{^{+}}_0\rangle \langle \hat{\phi}^{^{+}}_0}
{\hat{\lambda} - \hat{\lambda_0}} + \hat{{\cal K}}_{++}
\]
\[
P_+\hat{\Lambda}P_- = \frac{\hat{\phi}^{^{+}}_0\rangle \langle
\hat{\phi}^{^{-}}_0} {\hat{\lambda} - \hat{\lambda_0}} +
\sum_{s\neq 0} \frac{\hat{\phi}^{^{+}}_s\rangle \langle
\hat{\phi}^{^{-}}_s} {\hat{\lambda} - \hat{\lambda_s}} :=
 \frac{\hat{\phi}^{^{+}}_0\rangle \langle \hat{\phi}^{^{-}}_0}
{\hat{\lambda} - \hat{\lambda_0}} + \hat{{\cal K}}_{+-}
\]
\[
P_-\hat{\Lambda}P_+ =
 \frac{\hat{\phi}^{^{-}}_0\rangle \langle \hat{\phi}^{^{+}}_0}
{\hat{\lambda} - \hat{\lambda_0}} + \hat{{\cal K}}_{-+},\,
\hat{{\cal K}}_{-+} = \left( \hat{{\cal K}}_{+ -} \right)^{^+}
\]
\[
P_-\hat{\Lambda}P_- = \frac{\hat{\phi}^{^{-}}_0\rangle \langle
\hat{\phi}^{^{-}}_0} {\hat{\lambda} - \hat{\lambda_0}} +
\sum_{s\neq 0} \frac{\hat{\phi}^{^{-}}_s\rangle \langle
\hat{\phi}^{^{-}}_s} {\hat{\lambda} - \hat{\lambda_s}} :=
 \frac{\hat{\phi}^{^{-}}_0\rangle \langle \hat{\phi}^{^{-}}_0}
{\hat{\lambda} - \hat{\lambda_0}} + \hat{{\cal K}}_{--}.
\]
Then non-dimensional expression  (\ref{DNr}) may be  presented  as
\begin{equation}
\label{DNapprox} R{\Lambda}^r = \hat{\Lambda}^r =
\frac{\hat{\phi}^{^{+}}_0\rangle \langle \hat{\phi}^{^{+}}_0}
{\hat{\lambda} - \hat{\lambda_0}} + \hat{{\cal K}}_{+-} \,\,-
\left[\frac{\hat{\phi}^{^{+}}_0\rangle \langle
\hat{\phi}^{^{-}}_0} {\hat{\lambda} - \hat{\lambda_0}} +
\hat{{\cal K}}_{+-}\right] \frac{I}{
\frac{\hat{\phi}^{^{-}}_0\rangle \langle \hat{\phi}^{^{-}}_0}
{\hat{\lambda} - \hat{\lambda_0}} + \hat{{\cal K}}_{--} +
\sqrt{\frac{m^{\parallel}}{m^{\bot}}}R K^{^{-}}}
 \left[\frac{\hat{\phi}^{^{-}}_0\rangle \langle \hat{\phi}^{^{+}}_0}
{\hat{\lambda} - \hat{\lambda_0}} + \hat{{\cal K}}_{-+}\right].
\end{equation}
The  operator
 $
 R K^{-} = \hat{{K}}^{-}$
 for selected value of  energy is  positive and  may  be  estimated
  from below by the non-dimensional distance from the Fermi-level to the  second
  threshold with (non-dimensional) coefficient $\frac{R}{\delta}$:
  \[
\hat{K}^{-} \geq \sqrt{\frac{3 m^{^{\parallel}}}{2 m^{\bot}}}
\frac{R}{\delta} I_{-}
  \]
In particular, estimating  the contribution  ${\cal K}$ from the
 non-resonance terms  to  DN-map as $\parallel\hat{\cal K}\parallel
 \leq \frac{\hat{C}}{\hat{\rho(\hat{\lambda}_0)}}$, we may
  conclude that
  \begin{equation}
\label{small}
\parallel \,\, \sqrt{\frac{m_{0}}{m^{\parallel}}}
\left[ \hat{K}^{-} \right]^{^{-1}} \hat{\cal K}\parallel \leq
\sqrt{\frac{2 m^{\bot}}{3 m^{\parallel}}} \frac{{\hat{C}}}{ \,\pi
\,\hat{\rho}(\hat{\lambda}_0) }\,\,\,\, \,\frac{\delta}{R}.
  \end{equation}
Assuming that $m^{\bot}= 0.190\ \ m_0,\,\, m^{\parallel}= 0.916\,
m_0,$ (for Si) $\,\,\hat{C}= 10, \,\,\hat{\rho}(\hat{\lambda}_0)=
2.3, \delta = 2 nm,\,\, R = 10 nm $, we  obtain in the  right side
of (\ref{small}) a ``small'' non-dimensional  parameter
 $\epsilon \approx  0.1$ which may be used  when
developing a perturbation procedure. In particular  we  may
calculate the  inverse of $
\left[\frac{\hat{\phi}^{^{-}}_{_0}\rangle \langle
\hat{\phi}^{^{-}}_{_{0}}} {\hat{\lambda} - \hat{\lambda_0}} +
\hat{{\cal K}}_{--} + \hat{K}^{^{-}}\right] $ estimating
$\left(\hat{\cal K}_{--} + \hat{K}^{-}\right)^{-1}: = k(\delta, R)
:= k $ as
\[
  \displaystyle
 \parallel k \parallel\leq \frac{\delta}{R}\,\,\frac{1}{\pi}
\sqrt{\frac{2 m^{\bot}}{3 m^{\parallel}}} ,\,\,\,\mbox{or}\,\,\,
\parallel k_{\delta, R}\parallel \leq 0.14\,\,\frac{\delta}{R}
\]
with above  assumption. Really, solving the  equation
\[
\left[ \frac{\hat{\phi}^{^{-}}_0\rangle \langle
\hat{\phi}^{^{-}}_0} {\hat{\lambda} - \hat{\lambda_0}} +
\hat{{\cal K}}_{--} + K^{^{-}} \right] u = f,
\]
we  obtain an  explicit expression for the  inverse  operator
\[
u = \left[\frac{\hat{\phi}^{^{-}}_{_0}\rangle \langle
\hat{\phi}^{^{-}}_{_{0}}} {\hat{\lambda} - \hat{\lambda_0}} +
\hat{{\cal K}}_{--} + \hat{K}^{^{-}}\right] ^{^{-1}} f = k\,\,f -
\frac{1}{\cal D}\,\,\,k\, \,
\hat{\phi}^{^{-}}_{0}\rangle\,\,\langle \hat{\phi}^{^{-}}_{0},\,
k\,\, f \rangle,
\]
where ${\cal D} = \hat{\lambda} - \hat{\lambda}_0 + \langle
\hat{\phi}^{^{-}}_{0},\, k\,\, \, \hat{\phi}^{^{-}}_{0}\rangle$.
Substituting that expression into (\ref{DNapprox}), we  notice
that  all terms containing  $\left(\hat{\lambda}-
\hat{\lambda_0}\right)$ in denominator are cancelled and  we
obtain: $$  \hat{\Lambda}^r =
 \frac{\hat{\phi}_{_{0}}^{^r} \rangle\,\,
 \langle \hat{\phi}^{^r}_{_{0}}}{\hat{\lambda} - \hat{\lambda}_{_0}^{^r}}
 = \frac{\hat{\phi}_{_{0}}^{^+} \rangle\,\,
 \langle \hat{\phi}_{_{0}}^{^+}}{\cal D}, $$
with the  residue  of  the  DN- map
 proportional to  $\hat{\phi}_{_0}^{^+}\rangle \,\,
 \langle \hat{\phi}_{_0}^{^+}$, as  announced.
The  eigenvalues $\hat{\lambda}^{^r}$ of  the  operator
$\hat{l}^{^r}$ can be obtained from the  equation  ${\cal
D}(\hat{\lambda}_r) = 0 $. The  first  order correction gives due
to (\ref{smallpar}) $ |\hat{\lambda}^{^r}_{_0} -
\hat{\lambda}_{_0}|  <
\frac{||\hat{\phi}_{_{0}}^{^-}||^{^{2}}}{44} \approx 0.023 < < 2.3
$,
 in agreement with the  corresponding
 direct calculation  \cite{Olejnik}.


\vskip0.5cm


\begin{thebibliography}{99}

\bibitem{Adamyan}  V. Adamyan. \textit{Scattering matrices for
microschemes}.Operator Theory: Adv. and Appl. \textbf{59} (1992).


\bibitem{Alamo}J. del Alamo,\, C. Eugster,
 Appl. Phys. Letters {\bf 56}(1),\, January 1,\,(1990).

\bibitem{Albeverio}  S. Albeverio, F. Gesztesy, R. Hoegh-Krohn, H. Holden,
\textit{Solvable models in quantum mechanics} (Springer-Verlag,
New York 1988).

\bibitem{Kurasov}  S.Albeverio, P.Kurasov \textit{Singular Perturbations of
Differential Operators},( London Math. Society Lecture Note Series
271. Cambridge University Press 2000)

\bibitem{Helsinki2} N.T.Bagraev, A.B.Mikhailova, B.S.Pavlov, L.V.
Prokhorov, A.M.Yafyasov in 10-th MEL-ARI/NID Workshop,Helsinki,1-3
July, (Book of Abstracts,Academy of Finland, 2002)

\bibitem{B1}  N.T. Bagraev, V.K. Ivanov, L.E. Kljachkin, A.M. Malyarenko,
I.A. Shelych, Semiconductors ,  {\bf 34}, 6 (2000).

\bibitem{B2}  N.T. Bagraev, V.K. Ivanov, L.E. Kljachkin, A.M. Maljarenko,
S.A. Rykov, I.A. Shelyh, Semiconductors, {\bf 34}, 7 (2000).

\bibitem{B3}  N.T. Bagraev, A.D. Bouravleuv, A.M. Malyarenko, S.A. Rykov,
 Low-Dim. Structures.{\bf 9/10} (2000).

\bibitem{Boston} N. Bagraev,B. Pavlov, A. Yafyasov in: {\it International Conference on
Simulation of Semiconductor Processes and Devices, 3-6  September
2003} ( MTI,Book of Abstracts, Boston,2000)  .

\bibitem{MP00}  V. Bogevolnov, A, Mikhailova, B. Pavlov, A. Yafyasov In:
{\it Operator Theory : Advances and Applications", Vol 124 }
 Ed. A.Dijksma, A.M. Kashoek, A.C.M.Ran ( Birkh\"{a}user, Basel 2001)

\bibitem{Buttiker85} M.B\"{u}ttiker,Y.Imry,R.Landauer,S.Pinhas,
 Phys.Rew.B {\bf 31}(1985).

\bibitem{Buttiker93} M.B\"{u}ttiker, H.Thomas, A.Pretre,
 Phys. Rev. Lett. {\bf 70} (1993) .

\bibitem{ByRashba84} Y.A. Bychkov, E.I.Rashba,  J. Phys. C {\bf
17}, 6039 (1984).

\bibitem{Compano}  European Comission IST programme: Future and emerging
Technologies. \textit{Technology Roadmap for Nanoelectronics},Ed.
by R. Compano,Second edition (Luxemburg: Office for Official
Publications of the European Communities, 2000).

\bibitem{Cohen03} D. Cohen, Phys. Rev. B {\bf 68},
201303, (2003)

\bibitem{CsXu02} D.Csontos H.Q.Xu,  J. Phys.: Condensed
Matter {\bf 14} (2002), 12513

\bibitem{Exner88}  P. Exner, P. \^{S}eba, Phys. Lett. A 129:8,9 (1988).

\bibitem{Exner96} P. Exner,\,P. \^{S}eba ,\,M.Tater,
Journal of  Mathematical Physics,{\bf 37}, 10,(1996).

\bibitem{Opening}  M. Faddeev, B. Pavlov, Proc. LOMI, v126 (1983). (English
Translation J. of Sov. Math. v27 (1984))

\bibitem{GP}  N. Gerasimenko, B. Pavlov, Theor. and Math. Phys. {\bf 74} (1988).

\bibitem{Gohberg} I.S. Gohberg and E.I. Sigal.
Mat. Sbornik. Acad. of  Sciences URSS, 84(1971).

\bibitem{Interf2}H.Ishio,K.Nakamura, J.Phys. Society  Japan{\bf 61}
(1992).

\bibitem{MPP02}  A. Mikhailova, B. Pavlov, I. Popov, T. Rudakova, A. Yafyasov,
 Mathematishe  Nachrichten, {\bf 235},1 (2002).

\bibitem{Interf1} H. Kasai,K. Mitsutake, A. Okiji,
J.Phys. Society Japan{\bf 60} (1991).

\bibitem{Schrader}  V. Kostrykin and R. Schrader, J. Phys. A: Math. Gen. v32 (1999).

\bibitem{KZ02}P.Kuchment, H. Zengh, Waves Random Media {\bf 12} (2002).

\bibitem{KZ01}P. Kuchment, H. Zengh, Journal  Math. Anal.
Appl. {\bf 258} (2001), pp 671-700

\bibitem{Landauer70} R.Landauer, Phil. Mag. {\bf 21}(1970) pp
863-867

\bibitem{Lax} P. Lax, R. Phillips {\it Scattering theory} (Academic
press,New York,1967)

\bibitem{Shur}  {\it Handbook series on semiconductor parameters} Ed. by
M.Levinshtein S. Rumyantsev and M. Shur (Word Scientific,
Singapore, New Jersey, London, 1999).

\bibitem{Embedding} V. M. Maz'ja. {\it Sobolev Spaces,} (Springer Series in Soviet
Mathematics. Springer-Verlag, Berlin ,1985).

\bibitem{MP01}  A.Mikhailova, B. Pavlov In: Unconventional models of Computations UMC'2K, ed. I.
Antoniou, C. Calude, M.J. Dinneeen (Springer Verlag series for
Discrete Mathematics and Theoretical Computer Science 2001).

\bibitem{MP02} A. Mikhailova,B.Pavlov  In: {\it Operator Theory : Advances
and Applications 132 : Operator Methods in Ordinary and Partial
Differential Equations} (S.Kovalevski Symposium, Univ. of
Stockholm, June 2000), Ed. S.Albeverio, N.Elander, W.N.Everitt and
P.Kurasov (Birkh"user, Basel-Boston-Berlin, 2002 )

\bibitem{QSW} A. Mikhailova, B. Pavlov, L. Prokhorov,
xxx.lanl.gov/arXiv/archive math-ph/0312038 

\bibitem{MuzAd} B.A.Muzykantskii, Y. Adamov, Physical Rviews B {\bf 68},155304 (2003)

\bibitem{Novikov}  S.P. Novikov in  {\it The Arnol'dfest (Proceedings of the
Fields Institute Conference in Honour of the 60th Birthday of
Vladimir I. Arnol'd) }, eds. E. Bierstone, B. Khesin, A.
Khovanskii and J. Marsden,(Communications of Fields Institute,
AMS, 1999).

\bibitem{Olejnik} The  estimation of the  shift  of the  resonance
eigenvalue was done  by  V. Olejnik.

\bibitem{PalmThil92} T.Palm, L.Thylen  Applied Phys. Letters {\bf 60}
(1992)

\bibitem{P00}  B.Pavlov. Provisional Patent: \textit{A System and Method for
Resonance manipulation of Quantum Currents Through Splitting}
(Auckland University Limited, 504590,New Zealand,17 May 2000) {\it
"Quantum Domain relay" },United  States Patent  application
10/276,952,\,(Patent. Appl. Publication US 2003/0156781 A1,  Aug.
21, 2003).

\bibitem{P01}  B.Pavlov in {\it 7-th MELARI/NID Workshop, Bellaterra, Barcelona,
February 7-9 ,2001} (Book of Abstracts,2001).

\bibitem{DN01} B.Pavlov In: {\it Scattering (Encyclopedia of Scattering)}, ed.
R. Pike, P. Sabatier, Academic Press (Harcourt Science and  Tech.
Company, 2001).

\bibitem{P02} B.Pavlov in:
{\it Proceedings  of  the  Center of  Mathematics and
Applications}, {\bf 40}, ( Ed. by Australian National University
2002).

\bibitem{Extensions}  B. Pavlov, Russian Math. Surveys 42:6 (1987).

\bibitem{Adil1} Radantsev V.F., Yafyasov A.M., Bogevolnov Semicond.Sci.Technol.16 (2001).

\bibitem{Rashba} E.I.Rashba  ZhETF {\bf 52} (1968).

\bibitem{Robert} Calculation of  components of the resonance entrance
vector was done  Mr. Kieran Robert, Department of Math. the
University of Auckland, NZ.

\bibitem{RS01} J.Rubinstein,M.Shatzman, Arch. Ration. Mech.
Analysis {\bf 160},4 (2001).

\bibitem{Safi99} I. Safi The European  Physical Journal B {\bf 12}
(1999).

\bibitem{Schat96} M. Schatzman
Applicable Anal. {\bf 61} (1996).

\bibitem{Samuelson}  L. Samuelson in:
7th MELARI/NID Workshop, Bellaterra, Barcelona, February 7-9
(2001).( Book of abstracts,2001).

\bibitem{Sarma} S.Das. Sarma, J. Fabian, X.Hu and I.Zutich. Solid
State Communic. {\bf 119},207(2001)

\bibitem{SU2}  J.Sylvester,\thinspace G. Uhlmann  in: Proceedings of the Conference
`` Inverse problems in partial differential equations
(Arcata,1989)'' (SIAM, Philadelphia, 1990).

\bibitem{Carbon01}C.Thelander,M.Magnusson,K.Deppert,L.Samuelson
Applied Physics letters {\bf 79} 13 (2001).

\bibitem{Esaki}  R.Tsu, L. Esaki, Applied Phys.
Letters., {\bf 22},562 (1973)

\bibitem{Kouw2002}W.G.Van der Wiel, Yu.V. Nazarov, S. De
Franceschi, T. Fujisawa, J.W. Elzerman, E.W.G.Huizeling, S.Tarucha
and L.P. Kouwenhoven ,Phys. Rev. B, November 2002.

\bibitem{Adil2} A.Yafyasov, V.Bogevolnov, A.Perepelkin
Phys.Stat.Sol.(b),183,(1994).

\bibitem{Aver_Xu01} H.Q.Xu, Applied  Phys. Letters {\bf
78} (2001), 2064

\bibitem{ScattXu02} H.Q.Xu, Applied Phys. Letters {\bf
80} (2002),853


\end{thebibliography}
\end{document}